\documentclass{jfm_modified_2021}
\usepackage{graphicx}
\usepackage{amsmath, amssymb}
\usepackage[inactive,tightpage]{preview}
\usepackage{import}
\usepackage{microtype}
\usepackage[normalem]{ulem}
\usepackage{mathtools,bm}
\usepackage{cases}
\usepackage{esvect}
\usepackage{enumitem}

\pdfmapfile{+rsfso.map}
\DeclareSymbolFont{rsfso}{U}{rsfso}{m}{n}
\DeclareSymbolFontAlphabet{\mathscr}{rsfso}

\usepackage{pdflscape}
\usepackage{afterpage}
\usepackage{flafter}
\usepackage{rotating}
\usepackage{fancyhdr}
\fancypagestyle{lscape}{%
\fancyhf{} 
\fancyfoot[LE]{}
\fancyfoot[LO] {}
 
}

\usepackage{bm}
\usepackage{microtype}

\usepackage{hyperref}
\usepackage[usenames, dvipsnames]{xcolor}
\hypersetup{colorlinks=true,linkcolor=MidnightBlue,citecolor=MidnightBlue}
\usepackage{natbib}
\usepackage{array}
\usepackage{booktabs}
\usepackage{multirow}
\usepackage{siunitx}
\usepackage{mathtools}
\usepackage{cleveref}

\usepackage{caption}
\usepackage{subcaption}
\usepackage{tabularx}
\newcolumntype{Y}{>{\centering\arraybackslash}X}

\usepackage{tikz}
\usepackage{pgfplots}
\usetikzlibrary{backgrounds, calc, patterns.meta, shadows, arrows, patterns}

\usetikzlibrary{arrows, decorations.markings}

\usepackage[outline]{contour} 
\contourlength{0.15em}

\newcommand{\edit}[1]{#1}
\newcommand{\editt}[1]{#1}

\usetikzlibrary{external}

\usepackage{grffile}
\pgfplotsset{compat=newest}
\usetikzlibrary{plotmarks}
\usetikzlibrary{arrows.meta}
\usepgfplotslibrary{patchplots}

\usepackage[outline]{contour} 
\contourlength{0.15em}

\usepackage{lscape}

\newcommand*{\ep}{\epsilon}

\newcommand*{\Oh}{\mathcal{O}}

\shorttitle{Asymptotic analysis of catchment models}
\shortauthor{Morawiecki and Trinh}

\title{On the development and analysis of coupled surface-subsurface models of catchments. \\[0.3em] \large Part 3. Analytical solutions and scaling laws}

\author{Piotr Morawiecki\corresp{\email{piotr.morawiecki@bath.edu}}
 \and Philippe H. Trinh\corresp{\email{p.trinh@bath.ac.uk}}}

\affiliation{
    Department of Mathematical Sciences, University of Bath, Bath BA2 7AY, UK
}

\pubyear{XXXX}
\volume{YYY}
\pagerange{xxx--xxx}
\date{\today~[Draft]}

\def\d{\mathrm{d}}
\newcommand*{\de}{\operatorname{d\!}{}} 
\newcommand{\dd}[2]{\dfrac{\de#1}{\de#2}}

\newcommand{\pd}[2]{\frac{\partial#1}{\partial#2}}
\newcommand{\dx}[1]{\frac{\partial #1}{\partial x}}
\newcommand{\ddx}[1]{\frac{\partial^2 #1}{\partial x^2}}

\newcommand{\dxp}[1]{\frac{\partial #1}{\partial x'}}

\newcommand{\dz}[1]{\frac{\partial #1}{\partial z}}

\newcommand{\dt}[1]{\frac{\partial #1}{\partial t}}
\newcommand{\dT}[1]{\frac{\partial #1}{\partial T}}

\newcommand{\dzhat}[1]{\frac{\partial #1}{\partial \hat{z}}}

\newcommand{\Pe}[0]{\mathrm{Pe}}
\newcommand{\tsat}[0]{t_\mathrm{crit}}
\newcommand{\Tsat}[0]{T_\mathrm{crit}}
\newcommand{\qsat}[0]{Q_\mathrm{crit}}

\begin{document}

\maketitle


\begin{abstract}
    \noindent The objective of this three-part work is to formulate and rigorously analyse a number of reduced mathematical models that are nevertheless capable of describing the hydrology at the scale of a river basin (\emph{i.e.} catchment). Coupled surface and subsurface flows are considered. 
    
    In this third part, we focus on the development of analytical solutions and scaling laws for a benchmark catchment model that models the river flow (runoff) generated during a single rainfall. We demonstrate that for catchments characterised by a shallow impenetrable bedrock, the shallow-water approximation allows a reduction of the governing formulation to a coupled system of one-dimensional time-dependent equations for the surface and subsurface flows. Asymptotic analysis is used to derive semi-analytical solutions for the model. We provide simple asymptotic scaling laws describing the peak flow formation\edit{, and demonstrate its accuracy through a comparison with the two-dimensional model developed in Part 2}. These scaling laws can be used as an analytical benchmark for assessing the validity of other physical, conceptual, or statistical models of catchments.
\end{abstract}


\section{Introduction}

\noindent In this third and final part of our work, we leverage the parametric study (Part 1, \citealt{paper1}) and two-dimensional benchmark models (Part 2, \citealt{paper2}) to perform an in-depth asymptotic analysis of a coupled surface-subsurface model of a catchment. \edit{We specifically focus on flow within an aquifer characterised by a thin porous layer. The system begins in a steady state, for a constant precipitation $r_0$, which is characterised by an initial seepage zone. Our objective is to understand the response of the catchment, when subjected to intense rainfall $r>r_0$.}

\begin{figure}
    \centering
    \includegraphics{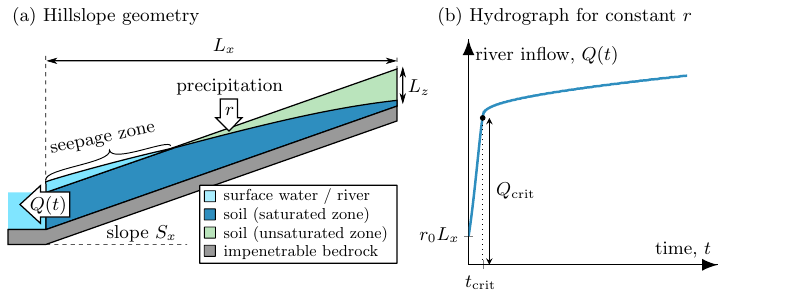}
    \caption{\edit{(a) Studied hillslope geometry; initially groundwater and surface water are in steady state for precipitation rate $r_0$, and therefore river inflow (per unit length) is $Q(0)=r_0L_x$. The simulated rise of river inflow caused by a constant rainfall $r>r_0$ is presented in (b). Note the characteristic fast rise of the river inflow to $\qsat$ at $t\in[0,\tsat]$.}}
    \label{fig:1D_model_introduction}
\end{figure}

\edit{One of the main conclusions from the numerical simulations in Part 2, is that in early time, the river inflow rapidly increases as a result of the rainfall accumulating over the initial seepage zone. It eventually reaches a \textit{critical flow}, followed by a much slower rise in the river inflow caused by the expansion of the seepage zone (see Fig. \ref{fig:1D_model_introduction}). One of the primary results we derive in this work, is an analytical solution for both early and late times, including a simple analytical formula for the critical flow:}
\begin{equation} \label{eq:intro_scale}
    \qsat = \underbrace{K_s S_x L_z}_{\text{groundwater flow}} + \underbrace{r L_x \left(1 - \frac{K_s S_x L_z}{r_0 L_x}\right)}_{\text{overland flow}}.
\end{equation}
This formula consists of a contribution from the groundwater flow, and a contribution from the overland flow formed over the seepage zone. The parameters $L_x$, $L_z$, and $S_x$ relate to geometrical features of the hillslope, $K_s$ corresponds to the soil hydraulic conductivity, while $r_0$ and $r$ represent the initial and simulated rainfall intensity, respectively.

We argue that such analytical scaling laws are valuable, both as a tool to diagnose the correctness of other, more complex rainfall-runoff models, and also as a measure for characterising a catchment's propensity to flooding.

The introduction of Part 1 covers the general subject of hydrology and parameter estimation, while the introduction of Part 2 covers computational rainfall-runoff models. We begin by discussing the content of this current paper, focusing on asymptotics and scaling laws, in the context of the existing literature.

\subsection{On the importance of analytical benchmarks results}

\noindent Catchment hydrology is one of many areas of engineering where numerical approaches tend to dominate over analytical ones. Due to the complex multiscale nature of hydrology, limited data availability and high computational cost, formulating and solving the correct equations can be enormously challenging~\citep{grayson1992physically}. Instead of the physical models based on the fundamental laws of hydrodynamics, simpler models such as conceptual and statistical models are often used instead~\citep{moore2007rainfall}. They are usually developed following a trial-and-error approach to fit available real-world data. However, these models do not provide any guarantee of model performance when applied to situations that may be underrepresented or missing in the training datasets~\citep{beven2018hypothesis, parkin1996validation, bathurst2004validation}.

To better understand the theoretical limits of different classes of catchment models, it is crucial to have a solid understanding of the different processes characterising the physical models. This goal can be achieved by developing numerical benchmark scenarios, as done by \emph{e.g.} by~\cite{sulis2010comparison} and~\cite{maxwell2014surface}, and using them to compare predictions of different models. One perspective is that, in order to ensure consistency between the models over a wide range of input parameters, models should predict the same scaling laws for key features. For example, if one model predicts that the peak river flow, $Q$, is proportional to the catchment area, $A$, so $Q\propto A$, while another model predicts that $Q\propto\sqrt{A}$, then regardless of the fitting of these models, they cannot give consistent predictions over the entire range of $A$ values. For example, the second model fitted to a training dataset dominated with measurements conducted in large catchments would tend to overestimate flow in the case of small, often ungauged, catchments. 

\edit{The above perspective may seem oversimplistic. However, we argue that despite many statistical works demonstrating such scaling laws [cf. \citealt{cunnane1987review,kjeldsen2008improving}] the comparison of asymptotic scaling laws between different catchment models is not a commonly used approach in the modern hydrology [though see \cite{vieira1983conditions} for comparison of Saint Venant approximations, and \cite{cook2009steady} for comparison of Richards and Boussinesq-based models].}

The emphasis of our work, here, on deriving of such analytical scaling laws for flow prediction in the case of coupled surface-subsurface flows in catchments. Not only do these scaling laws provide clear guidance on the key dependencies of model parameters, but their analysis often illuminates further model simplifications.

Next, we provide a brief overview of the existing analytical theory, highlighting the lack of research on fully-coupled surface-subsurface systems.

\subsection{On analytical solutions in catchment hydrology}

\noindent The typical governing equations used in physical catchment models include Richards equation for the subsurface flow (or the Boussinesq equation for the unconfined groundwater flow) and the Saint Venant equations for the overland and channel flows (see \emph{e.g.} the review by \citealt{shaw2015hydrology}). These equations and their simplifications have been well-studied using analytical methods, although largely in an uncoupled manner. \edit{One of the equations studied is the Boussinesq equation, which is commonly used to determine the shape of the groundwater table. In this paper by \textit{Boussinesq equation} we refer to the Boussinesq equation used to describe unconfined groundwater flow (see \emph{e.g.} \citealt{troch2013importance} and \citealt[chap.~2]{halek2011groundwater}), in order to distinguish it from other equations and approximations known by the same name.} In some cases, analytical solutions for this equation can be derived. Examples include for example steady-state groundwater flow and evolution in one-dimensional hillslopes \citep{polibarinova1962theory,troch2013importance}. Similarly, analytical solutions have been developed for the 1D Richards equation to describe water transfer through the unsaturated soil under constant and time-varying infiltration~\citep{warrick1990analytical}.

For the case of overland flow over a hillslope, analytical solutions have been found for a kinematic approximation of the Saint Venant equations, as done \emph{e.g.} by~\cite{parlange1981kinematic} and~\cite{tao2018approximate}. These analytical solutions and their approximations are important as they provide benchmark models for testing numerical schemes (\emph{e.g.} benchmarks by~\cite{macdonald1995comparison} for the overland flow and by~\cite{tracy2006clean} for the subsurface flow), and can be used to develop less computationally demanding modelling approaches such as TOPMODEL by~\cite{kirkby1979physically}. However, no analytical solutions have been found so far coupled systems that include governing equations describing both subsurface/groundwater and surface flow. Despite the importance of these models in catchment hydrology, the study of these models has been restricted to numerical solutions only~\citep{maxwell2014surface}.

\subsection{On the shallow water approximation for subsurface flow}

\noindent Previously, in Part 2, we introduced \textit{deep aquifer} scenario to describe a catchment with a deep aquifer, and \textit{shallow aquifer} scenario to describe a catchment where the subsurface flow is predominantly transferred through a thin porous layer near the surface. Mathematically, \textit{shallow aquifer} scenario is the limiting case of \textit{deep aquifer} scenario in which the aquifer depth is much smaller than the catchment width, $L_z\ll L_x$. We showed that in both cases, under standard initial conditions, the full three-dimensional catchment model can be reduced to a simpler two-dimensional hillslope model. We shall begin with this two-dimensional assumption as the basis in this paper. We then demonstrate that in the \textit{shallow water} scenario, under certain assumptions, the two-dimensional model can be further reduced to a one-dimensional model.

The reduction of the two-dimensional subsurface flow into a one-dimensional model is not a new concept. The simplification is based on the Dupuit-Forchheimer approximation by \cite{dupuit1863etudes} and \cite{forchheimer1914hydraulik}, which states that the groundwater flow is predominantly horizontal, and that the total flow scales proportionally with the saturated aquifer thickness. \cite{boussinesq1877essai} used this assumption to develop a one-dimensional model for the groundwater height; this is now known as the \editt{\emph{Boussinesq equation} for groundwater flow (see \textit{e.g.} \citealt{bartlett2018class}), or the \emph{Dupuit-Boussinesq equation} (see e.g. \citealt{guerin2014response})}. As we show later, it can be derived from the 2D Richards model under the aforementioned assumption that $L_z\ll L_x$. The accuracy of this approximation is studied in detail by~\cite{paniconi2003hillslope} and~\cite{cook2009steady}.

The Boussinesq equation is commonly used in groundwater modelling, and a wide class of analytical and approximate solutions has been developed. Notable examples are reviewed by~\cite{wooding1966groundwater}, \cite{anderson1996advances}, \cite{troch2003hillslope}, and~\cite{bartlett2018class}. These studies,  however, concern only groundwater flow, and do not involve the coupling with the overland flow, which is an essential component of a standard physical catchment model.

Here, we extend these studies by including the effect of overland flow in the Boussinesq equation. Our main result in this paper is the derivation and analysis of the following one-dimensional dimensionless coupled surface/subsurface model:
\begin{equation*}
    \dt{H} =
    \begin{cases}
        f(x)^{-1}\left[\dx{}\left(\sigma H\dx{H} + H\right) + \rho_0 r(x, t)\right] & \quad \text{if } H \leq 1, \\
        \dx{}\left(\sigma\dx{H} + \mu\left(H-1\right)^k\right) + \rho_0 r(x, t) & \quad \text{if } H > 1,
    \end{cases}
\end{equation*}
where $f(x)$, $\sigma$, $\mu$, and $\rho_0$ are dimensionless parameters explained in detail in \cref{sec:1D_formulation}, and $H(x,t)$ is the total height of groundwater and surface water\edit{, which depends on the distance from the channel $x$ and time $t$. Values $H\leq 1$ represent unsaturated soil without surface water, and $H>1$ represent saturated soil with surface water.} The main difference from the classical Boussinesq equation is the second case in the above equation with $H>1$, in which we include an additional term representing the overland flow.

We shall use the above one-dimensional coupled surface-subsurface model to develop analytical solutions for the river flow formed by rainfall of a constant intensity (however, the result can be generalised for time-dependent rainfall). Our methodology \edit{takes advantage of the negligibly small size of the diffusion terms in most of the seepage zone, which allows us to} use the method of characteristics for the study of wave propagation. This approach is similar to previous kinematic treatments of the Saint Venant equations \citep{woolhiser1967unsteady, henderson1964overland}, but for our problem, the size of the seepage zone increases as a result of rising groundwater, which introduces a secondary dynamics.

The analytical approximations we develop in this work are possible due to a few governing assumptions; they are supported by the analysis of the typical values of catchment parameters described in Part 1~\citep{paper1}. The main approximations are: (i) the typical rainfall duration is much shorter than the characteristic timescale of groundwater flow; (ii) the typical timescale of surface flow is much shorter than that of subsurface flow; and (iii) the mean precipitation rate is larger than the maximal groundwater flow passing through the saturated zone.

We start by introducing the above one-dimensional model in \cref{sec:1D_formulation}, with its typical dynamics discussed in \cref{sec:typical_dynamics}. For a scenario of single intensive rainfall described in \cref{sec:asymptotic_analysis_high_rho0}, we find an approximated analytical form of the initial steady state in \cref{sec:steady_state}, followed by a short-time asymptotic analysis in \cref{sec:short_time_asymptotics}. The accuracy of the developed analytical approximations is assessed in \cref{sec:numerical_comparison}. In \cref{sec:hydrograph_features}, we highlight key hydrograph features predicted by this analytical solution, followed by conclusions in \cref{sec:conclusions} and further discussion in \cref{sec:discussion}.

\section{Formulation of the one-dimensional coupled model}
\label{sec:1D_formulation}

\noindent In this section, we introduce a one-dimensional model describing the horizontal groundwater and overland flow along the hillslope, firstly in a dimensional and then in a dimensionless form. Its formal mathematical derivation from the two-dimensional benchmark model introduced in our previous paper is presented in~\cref{app:Boussinesq_derivation} --- here we focus on presenting the general structure of the model instead.

\subsection{Dimensional model}
\label{sec:model_formulation}

\begin{figure}
    \centering
    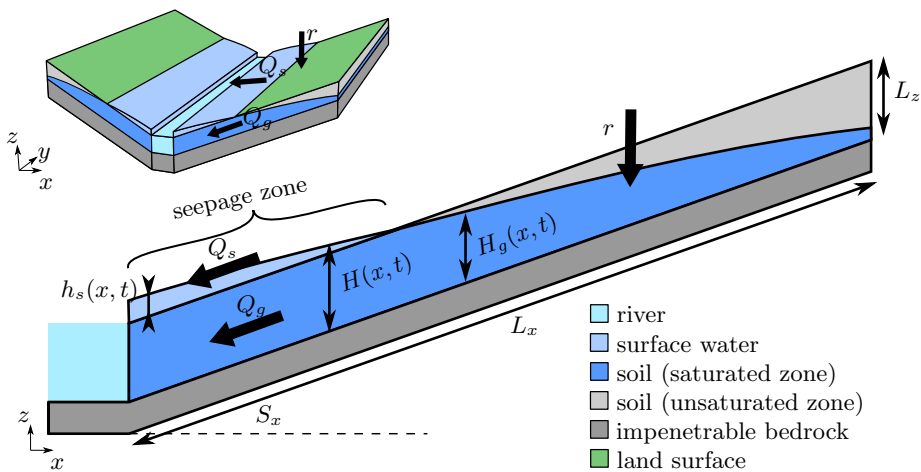
    \caption{Hillslope geometry used to formulate a 1D surface-subsurface model.}
    \label{fig:1D_hillslope}
\end{figure}

\noindent Let us consider a two-dimensional hillslope of length $L_x$ with a uniform terrain slope $S_x$, uniform thickness of the porous layer $L_z$, uniform saturated soil hydraulic conductivity $K_s$, and an impenetrable bedrock beneath the hillslope. As shown in the $(x,z)$-plane in \cref{fig:1D_hillslope}, we denote the thickness of the saturated zone as $H_g(x,t)$ and the height of the surface water as $h_s(x,t)$. 

We shall assume that overland flow can only occur when the soil becomes fully saturated. The overland flow generated by exceeding the soil infiltration capacity~\citep{kirkby2019infiltration} is not considered. Under this assumption, the heights $H_g$ and $h_s$ can be combined to form a single dependent variable, 
\begin{equation}
H(x,t) = H_g(x,t) + h_s(x,t),
\end{equation}
defined as the total height of groundwater and surface water. We now review the governing equations for $H$, for which the details are presented in \cref{app:Boussinesq_derivation}. 

\subsubsection{Groundwater flow}

\noindent The standard approach to model shallow-water aquifers uses the Dupuit–Forchheimer assumption, which states that groundwater flows horizontally with the pressure head following a hydrostatic profile. Under this approximation, the groundwater flow is given by
\begin{equation}
    \label{eq:dimensional_groundwater_flow}
    Q_g = K_s H\dx{H} + K_s H S_x.
\end{equation}
When the soil is not fully saturated, and hence $H<L_z$, the evolution of the groundwater depth is given by the continuity equation, resulting in a standard form of the Boussinesq equation~\citep{troch2013importance} for an unconfined aquifer:
\begin{equation}
    \label{eq:groundwater_dimensional}
    f(x, t) \dt{H_g} = \dx{}Q_g + r = \dx{}\left(K_s H_g\dx{H_g} + K_s H_g S_x\right) + r(x,t),
\end{equation}
where $r(x,t)$ denotes the groundwater recharge, and $f(x, t)$ is a drainable porosity. In this paper, we assumed that the recharge is equal to the precipitation. 

\edit{The drainable porosity function is more subtle; as introduced in \cref{app:drainable_porosity}, it is formally defined as the rate of change of groundwater volume, $\mathcal{V}$, given a change in the groundwater level, $H$, \emph{i.e.} $f=\de\mathcal{V}/\de H$. Note that $f$ depends on the soil saturation above the groundwater table. For example, higher soil saturation implies that less water is required to raise the groundwater by a given volume, and hence $f$ is lower. Recall that the soil saturation, $\theta$, is computed as a function of the pressure head, $h_g$, as given by the Mualem van-Genuchten model \eqref{eq:GMtheta}. In theory, computing $f$ would involve coupling equation \eqref{eq:groundwater_dimensional} with a model for $h_g(x, z,t)$.}

\edit{In the literature (\emph{e.g.}~\citealt{troch2003hillslope}), $f$ is often assumed to be a parameter with a value specific to the soil type at a given location. In practice, however, $f$ can change over time. For example, during a rainfall, a characteristic wetting front forms, which slowly propagates from the surface towards the groundwater table \citep{caputo2008front}. In order for this table to rise, the rainwater must first infiltrate through the unsaturated zone. This causes $f$ to significantly change over time.}

\edit{In this paper, we approximate $f(x, t)$ by a time-independent mean drainable porosity $f_{\text{mean}}(x)$. Although the model results will not duplicate the full time-dependent behaviour observed in Part 2, the mean value, $f_{\text{mean}}(x)$, is chosen such that solutions correctly capture the key time when the groundwater reaches the land surface. We shall see later that the resultant model reproduces the hydrograph during a single precipitation event (see later \cref{fig:hydrograph_approximations}).}

In \cref{app:drainable_porosity}, we justify the choice of 
\begin{equation}
    f(x, t) \approx f_{\text{mean}}(x) \equiv \frac{v_H(x)}{D(x)},
    \label{eq:mean_porosity}
\end{equation}
where $v_H(x)$ is the initial drainable volume per unit area at a given location, and $D(x)=L_z-H(x)$ is the depth of the groundwater below the land surface. In other words, the drainable porosity is given by the fraction of the soil volume that can be filled with water. Henceforth, we take the mean approximation \eqref{eq:mean_porosity} as the definition of $f$. Further discussion of the drainable porosity function is provided in~\cref{app:drainable_porosity}, where we provide formulae for its implementation.


\subsubsection{Coupling with the overland flow}
\label{sec:overland_coupling}

\noindent Now let's consider the case where the soil is fully saturated, for which $H_g = L_z$, which implies $H>L_z$. In this case, as derived in \cref{app:Boussinesq_derivation}, the surface depth $h_s$ evolves according to the following continuity equation:
\begin{equation}
    \label{eq:overland_dimensional}
    \dt{h_s} = \dx{}\left(Q_g + Q_s\right) + r = \dx{}\left(K_s L_z \dx{h_s} + \frac{\sqrt{S_x}}{n_s} h_s^k\right) + r(x,t).
\end{equation}
In the above equation, we have used Manning's equation to represent the overland flow:
\begin{equation}
    \label{eq:dimensional_Mannings}
    Q_s = \frac{\sqrt{S_f}}{n_s} h_s^k \sim \frac{\sqrt{S_x}}{n_s} h_s^k,
\end{equation}
where $k=5/3$ and $n_s$ is the Manning roughness coefficient, which depends on the hillslope surface type and is determined empirically. \edit{We use the kinematic approximation ($S_f \sim S_x$), where the friction slope $S_f$ is only dependent on the elevation gradient $S_x$. Alternatively, we could also consider the diffusive approximation $S_f\sim S_x+\dx{h_s}$; however we limit this study to the kinematic approximation only for simplicity.}

\edit{Furthermore, we note that apart from the constant gravitationally-induced groundwater flow, the pressure difference caused by the gradient of surface water height may affect the groundwater flow. Typically, the size of the overland flow is negligibly small compared to the thickness of the porous layer. However, as we shall discuss in \cref{sec:steady_state}, this approximation fails at the propagating seepage front, which requires us to include the $\dx{}\left(K_s L_z \dx{h_s}\right)$ diffusion term.}

Now, we can combine eqs~\eqref{eq:groundwater_dimensional} and~\eqref{eq:overland_dimensional} into a single equation for $H$:
\begin{equation}
    \label{eq:dHdt_dimensional}
    \dt{H} =
    \begin{cases}
        f(x)^{-1}\left[\dx{}\left(K_s H\dx{H} + K_s H S_x\right) + r\right] & \quad \text{if } H \leq L_z, \\
        \dx{}\left[K_s L_z \dx{H} + \frac{\sqrt{S_x}}{n_s} \left(H-L_z\right)^k\right] + r & \quad \text{if } H > L_z.
    \end{cases}
\end{equation}

We assume a no-flow boundary condition at the catchment boundary ($x=L_x$). At the location of the river ($x=0$), we assume that the river table is located \edit{at the same level as the overland water height (or at the surface if no overland flow is present)}. Therefore, at $x=0$ we set $H_g=L_z$ and a (flat) free-surface condition for the overland flow, $\partial h_s/\partial x = 0$. Additionally, in this work, we will study the time evolution of the above system assuming that it is initially in equilibrium for a given mean rainfall $r_0<r$, and then subjected to a rainfall $r$ for $t > 0$. Therefore, for the initial condition, we take the steady-state of equation of~\eqref{eq:dHdt_dimensional} for a given mean rainfall $r_0$. The boundary and initial conditions are summarised shortly in \cref{sec:nondim}.

\edit{\subsubsection{Channel flow}
\noindent As it was discussed in sec. 3.3 of Part 2, the total groundwater and overland flow that is reaching the riverbank form a channel flow. This flow can be described by one-dimensional Saint Venant equations. However, in this study, our focus is on studying the properties of river inflow from the hillslope, not the subsequent channel flow. Therefore for the purpose of this study, we assume that the river height is constant, limited to the depth of the channel. Analysing how the river inflow propagates thought the channel (or the entire drainage network) and how the drying of the channel impacts the surface and subsurface flows can be interesting extensions of this study.
}

\subsection{Non-dimensional model}

\noindent The above model can be nondimensionalised by taking $x = L_x x'$, $t = T_0 t'$, $H=L_z H'$ and $r=r_0r'$. Here, $T_0= L_x/(K_sS_x)$ is a characteristic timescale of the groundwater flow, chosen to balance the temporal term and the $\partial_x(K_s H S_x)$ term.

Once nondimensionalised, our governing equations~\eqref{eq:dHdt_dimensional} become (after dropping primes):
\begin{subequations} \label{eq:1Dmodel}
\begin{equation}
    \label{eq:dHdt_dimless}
    \dt{H} =
    \begin{dcases}
        f(x)^{-1}\left[\dx{}\left(\sigma H\dx{H} + H\right) + \rho_0 r(x, t)\right] & \quad \text{if } H \leq 1, \\
        \dx{}\left(\sigma\dx{H} + \mu \left(H-1\right)^k\right) + \rho_0 r(x,t) & \quad \text{if } H > 1,
    \end{dcases}
\end{equation}
and the dimensionless parameters $\sigma$, $\mu$, and $\rho_0$ are introduced shortly in~\cref{sec:nondim}. In this work, we assume that the rainfall is constant and uniform, \emph{i.e.} $r(x, t) = r = \text{constant}$, except for the initial jump from $r_0$ to $r>r_0$. However, we shall discuss in \cref{sec:discussion} that our methodology can be applied in the case of time-dependent rainfall.

\edit{In combination with the governing equations \eqref{eq:dHdt_dimless}, we have to specify the dimensionless boundary conditions. First, at $x = 0$, we need to consider two situations. Firstly, if a seepage zone exists for the initial $r_0$, we set a free flow boundary condition,
\begin{equation}
    \label{eq:bc_river_v1}
    H_x(0, t) = 0.
\end{equation}
However, as we shall demonstrate in \cref{sec:typical_results}, if $r_0$ is low enough, initially the seepage does not exist. Then we assume that $H(x,t)$ representing groundwater is reaching the surface at $x=0$, \emph{i.e.}:
\begin{equation}
    \label{eq:bc_river_v2}
    H(0, t) = 1.
\end{equation}
During a rainfall ($r>r_0$), the groundwater gradient at $x=0$, which is initially negative, increases as the groundwater rises. The seepage starts to grow, when the  when it becomes positive, which is when boundary condition \eqref{eq:bc_river_v2} is replaced with \eqref{eq:bc_river_v1}.}

At the right-hand edge, by definition of a catchment, there is zero flow:
\begin{equation}
	Q(1, t) = 0, \label{eq:no_flow_bc}	
\end{equation}
\end{subequations}
\noindent where the dimensionless total flow, $Q$, is defined as:
\begin{equation}
    \label{eq:q_dimless}
    Q(x, t) =
    \begin{dcases}
        H + \sigma H\dx{H} & \quad \text{if } H \leq 1, \\
        1 + \sigma\dx{H} + \mu(H-1)^k & \quad \text{if } H > 1.
    \end{dcases}
\end{equation}

\edit{For the initial condition, $H(x,t=0)=H_0(x)$, we take a steady state of \eqref{eq:dHdt_dimless} for $r=1$:
\begin{equation}
    \label{eq:H_initial_condition}
    0 =
    \begin{dcases}
        \dx{}\left(\sigma H_0\dx{H_0} + H_0\right) + \rho_0 & \quad \text{if } H_0 \leq 1, \\
        \dx{}\left(\sigma\dx{H_0} + \mu \left(H_0-1\right)^k\right) + \rho_0 & \quad \text{if } H_0 > 1,
    \end{dcases}
\end{equation}}
In this paper, we refer to this model \eqref{eq:1Dmodel} as the \textit{1D model}.

\begin{table}
    \centering
    \begin{tabular}{rcccc}
        \textsc{parameter} & \textsc{symbol} & \textsc{mean value} & \textsc{unit} \\
        Catchment width & $L_x$ & $616$ & $\mathrm{m}$ \\
        Catchment depth & $L_z$ & $1$ & $\mathrm{m}$ \\
        Hydraulic conductivity & $K_s$ & $10^{-4}$ & $\mathrm{ms^{-1}}$ \\
        Mean precipitation rate & $r_0$ & $2.95\cdot 10^{-8}$ & $\mathrm{ms^{-1}}$ \\
        Peak precipitation rate & $r$ & $2.36\cdot 10^{-7}$ & $\mathrm{ms^{-1}}$ \\
        Hillslope gradient & $S_x$ & $0.075$ & $-$ \\
        Manning's roughness coefficient & $n_s$ & $0.051$ & $\mathrm{sm^{-1/3}}$ 
    \end{tabular}
    \caption{Typical values of physical parameters characterising UK catchments extracted in Part 1 of this paper \citep{paper1}.
    }
    \label{tab:typical_parameters}
\end{table}

\subsection{The non-dimensional parameters} \label{sec:nondim}

\noindent In the first case of \eqref{eq:1Dmodel}, we have introduced two key dimensionless parameters, defined as
\begin{subequations}
\begin{align}
    \sigma &=\frac{L_z}{L_xS_x} = \frac{\textrm{thickness of the porous layer}}{\textrm{elevation drop along the hillslope}},\label{eq:sigma_def}\\
    \rho_0 &=\frac{r_0 L_x}{L_z S_x K_s} = \frac{\textrm{precipitation \editt{flux}}}{\textrm{\editt{maximum} groundwater flux}}, \label{eq:rho0_def}
\end{align}
\end{subequations}
Note that $\sigma \to \infty$ as the hillslope becomes increasingly flat. The parameter $\rho_0$ represents the ratio of the total precipitation rate (given by $rL_x$ in $\mathrm{m}^2 \mathrm{s}^{-1}$) to the maximum possible groundwater flow for fully saturated soil (given by $L_z S_x K_s$ in $\mathrm{m}^2 \mathrm{s}^{-1}$). \editt{Introduction of the maximum groundwater discharge is a classic concept in hydrology, see \textit{e.g.} \cite{horton1936maximum}.}

\edit{In the second case of \eqref{eq:1Dmodel}, we have introduced an additional dimensionless parameter:
\begin{equation}
    \mu = \frac{L_z^{k-1}}{K_sS_x^{1/2}n_s}.
    \label{eq:mu_def}
\end{equation}}
\edit{We shall argue later in \cref{sec:method_of_characteristics} that the characteristic size of the overland flow scales as $L_s=\mu^{-1/k}L_z$. Following that section, we introduce a key dimensionless parameter to describe the dynamics in the seepage zone, namely the Péclet number:
\begin{equation} \label{eq:peclet}
    \Pe  = \frac{\mu^{1/k}}{\sigma}=\frac{\frac{\sqrt{S_x}}{n_s}L_s^{k}}{K_sL_z\frac{L_s}{L_x}}=\frac{\text{convective overland flow}}{\text{diffusive effect of the groundwater flow}}.
\end{equation}
In order to interpret $\Pe$, we note that the numerator represents the second term on the right-hand side of \eqref{eq:dHdt_dimensional} for $H>L_z$, representing convective effects. The denominator represents the size of the first term on the right-hand side of \eqref{eq:dHdt_dimensional} for $H>L_z$, representing diffusive effects.}

\edit{Based on \editt{median values} of physical parameters used in the above equations provided in \cref{tab:typical_parameters}, we \editt{have} $\sigma\approx10^{-2}$, $\rho_0\approx 1.5$, $\mu\approx10^7$ and $\Pe\approx 10^5$. Consequently, our work will primarily focus on the limits of $\mu, \, \Pe \to \infty$, where convection dominates diffusion.}


\section{Numerical methodology and typical dynamics}
\label{sec:typical_dynamics}

\noindent Recall that the solutions are essentially characterised by the overland- and groundwater heights, and a quadruplet of parameters:
\begin{equation}
H = H(x, t; \, \sigma, \, \mu, \, \rho_0, \, r),
\end{equation}
as well as the constant $k = 5/3$ appearing in Manning's formula. These solutions are found by solving the PDE~\eqref{eq:dHdt_dimless} subject to the boundary conditions~\eqref{eq:bc_river_v1}-\eqref{eq:no_flow_bc}. In the typical configuration, the flow transitions from overland ($H > 1$) to groundwater ($H \leq 1$) at a contact point $x = a(t)$, which is determined as part of the solution.

\afterpage{
    \centering
    \includegraphics[width=0.98\linewidth]{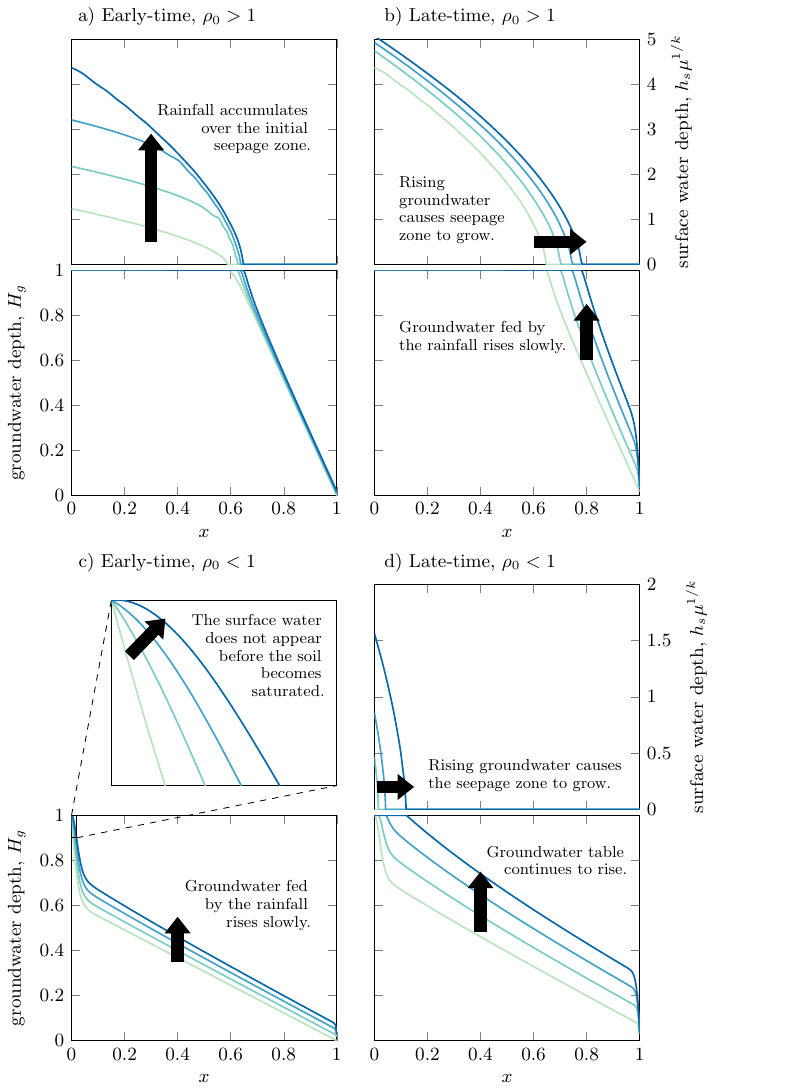}
    \captionsetup{type=figure}
    \captionof{figure}{Schematic presenting early- and late-time dynamics for a catchment with and without an initial seepage zone (corresponding to $\rho_0>1$ and $\rho_0\leq1$ respectively). \edit{In (a), (b) and (c) a seepage zone is observed, and therefore a separate graph is presented to illustrate the evolution of the surface water depth (note that the vertical axis is multiplied by the scaling factor $\mu^{1/k}\approx 3260$).} \label{fig:early_late_time_behaviour}}
    \clearpage
}

The model in~\eqref{eq:dHdt_dimless} was implemented in Matlab using the \texttt{ode15s} solver to find its steady state and \texttt{pdepe} to solve the time-dependent problem. We divide the spatial and temporal domain as follows:
\begin{equation}
x_i = \frac{i}{N_x} \quad \text{and} \quad t_j = \frac{j}{N_t} \, t_\mathrm{max}, \qquad i, j = 1, 2, 3, \ldots
\end{equation}
where we typically use $N_x = 200$ and $N_t = 300$. We check whether increasing mesh size and time resolution does not significantly impact the obtained solution, and if it does, we refine the mesh. The codes used to generate figures in this work are available in a GitHub repository~\citep{github2}. All numerical results in this paper were obtained for the values presented in \cref{tab:typical_parameters}, unless stated otherwise. 

\subsection{Typical results for \texorpdfstring{$\rho_0 < 1$}{} and \texorpdfstring{$\rho_0 > 1$}{}}
\label{sec:typical_results}

\noindent The existence of the seepage zone in the initial steady state depends on whether the value of $\rho_0$, defined in \eqref{eq:rho0_def}, is higher or lower than $1$, and each case exhibits a different transient behaviour. Typical solutions obtained in these two cases are shown in \cref{fig:early_late_time_behaviour}.

\begin{figure}
    \centering
    \includegraphics{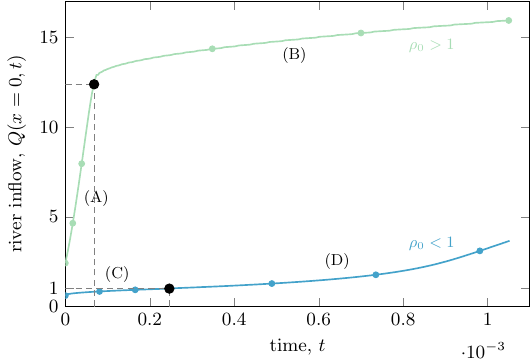}
    \caption{Schematic representation of the hydrograph obtained for a catchment with and without an initial seepage zone (corresponding to $\rho_0>1$ and $\rho_0\leq1$ respectively). Points represent times for which the profiles are shown in \cref{fig:early_late_time_behaviour}, while letters A-D refer to corresponding phases from that figure.}
    \label{fig:early_late_time_hydrograph}
\end{figure}

First, consider \cref{fig:early_late_time_behaviour}(a, b). In the case of $\rho_0>1$, we already have a seepage zone in the initial state. In a short timescale, the height of the surface water increases quickly until reaching a seemingly quasi-static state. Afterwards, the flow continues to increase as a result of the saturation front propagating uphill, but this process is characterised by a much longer timescale and a slower rate of flow rise. The difference between the short- and long-time behaviour can also be seen in the produced hydrograph in \cref{fig:early_late_time_hydrograph}. \edit{It shows the dependence between the total river inflow, defined as the total flow \eqref{eq:q_dimless} evaluated at the river bank ($x=0$),
\begin{equation}
    Q(t)=Q(x=0,t).
\end{equation}
In the presented hydrograph,} we have marked the initial fast transition as (A) and the subsequent slow transition as (B). 

For $\rho_0<1$, we do not observe an initial seepage zone, \emph{i.e.} $H(x,t=0)<1$ for all $x$. For some time the groundwater table is rising, increasing groundwater flow reaching the river, until the groundwater depth gradient at $x=0$ becomes $0$. Then the seepage zone starts to slowly form and propagate away from the channel, increasing the overland flow reaching the river.

However, in practice, in the case of real-world catchments characterised by a thin porous layer (which is a base assumption behind the presented 1D model), the groundwater flow rate is highly limited. Therefore, \edit{in such catchments, we expect the} $\rho_0>1$ case to be more prevalent, which is additionally confirmed by low BFI\footnote{BFI (Base Flow Index) describes the ratio between the base flow and total flow in the given catchment. High values refer to catchments dominated by the groundwater flow, while low values refer to catchments with a significant overland flow component.} values characterising low-productive catchments. Therefore, in this paper, we focus on discussing the mathematical properties of each phase presented in~\cref{fig:early_late_time_hydrograph} only in the case of $\rho_0>1$.

\afterpage{
    \centering
    \includegraphics{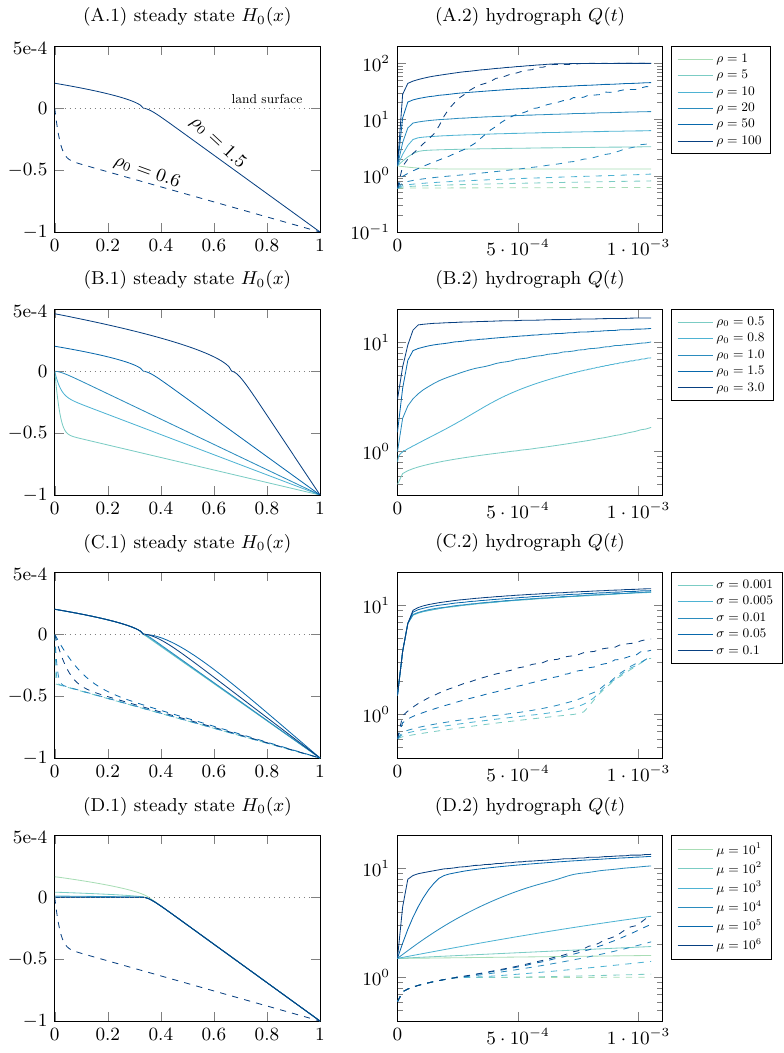}
    \captionsetup{type=figure}
    \captionof{figure}{Impact of dimensionless parameters shown for the initial steady-state $H_0(x)$ vs $x$ (insets left column) and for the hydrograph $Q(t)$ vs $t$ (insets right column). Insets (A) shows changing $\rho_0$; (B) for changing $\rho$; (C) for changing $\sigma$; and (D) for changing $\mu$. \edit{In the case of (A), (C), and (D), solid lines represent solutions for $\rho_0=1.5$ (scenario with an initial seepage zone), and dashed lines represent solutions for $\rho_0=0.6$ (scenario without an initial seepage zone). The surface water ($H>0$) is magnified 1000 times.}\label{fig:parameter_dependence}}
    \clearpage
}

In general, the solution for the PDE model~\eqref{eq:dHdt_dimless} can only be found numerically. However, by taking advantage of the typical sizes of dimensionless parameters, the model can be further simplified. \edit{Fig.~\ref{fig:parameter_dependence} shows the effect of dimensionless parameters, $\rho$, $\rho_0$, $\sigma$, and $\mu$ on the model's solution. The graphs in the left column show how the initial steady state $H_0(x)$ depends on the value of each parameter, while the graphs on the right present the impact of each parameter on the hydrograph $Q(x=0,t)$.} The conclusions from this numerical experiment are as follows:
\begin{enumerate}[label=(\roman*),leftmargin=*, align = left, labelsep=\parindent, topsep=3pt, itemsep=2pt,itemindent=0pt ]
\item \textbf{Parameter} $\mathbf{\rho_0}$ (typical value $2$)\edit{, which following \eqref{eq:rho0_def} characterises the mean precipitation rate in terms of the groundwater flow,} has a significant impact on the initial steady state. As discussed in detail before, $\rho_0<1$ corresponds to a hillslope with no initial seepage zone, and is characterised by different dynamics than the $\rho_0>1$ case, in which we observe a fast rise of flow in the early time. In most of this paper, we will consider only the latter case.
\item \textbf{Parameter} $\mathbf{\rho}$ (typical value $\approx 20$), \edit{which characterises the simulated precipitation rate in terms of the groundwater flow,} does not affect the initial steady state, but it does affect how quickly the flow is rising. Higher $\rho$ values lead to both a higher flow over the seepage zone and a faster growth of this zone. Values of $\rho$ can vary significantly depending on the rainfall event considered.
\item \textbf{Parameter} $\mathbf{\sigma}$ (typical value $10^{-2}$)\edit{, which following \eqref{eq:sigma_def} characterises the thickness of porous layer compared to the elevation drop along the hillslope,} has a significant impact only on the solution outside the seepage zone. In the case of the seepage zone, the term including $\sigma$ is negligibly small compared to the $\mu$ term. However, it affects the speed at which the seepage zone is growing. Note that as $\sigma\rightarrow 0$, the initial groundwater shape becomes a linear function (with a possible small boundary layer at its left border). Even though this limit is not critical in our analysis, it may allow us to approximate the initial groundwater shape without the need to solve the governing equations numerically.
\item \textbf{Parameter} $\mathbf{\mu}$ (typical value $10^6$)\edit{, which following \eqref{eq:mu_def} characterises the overland flux,} does not have a significant impact on the groundwater table outside the seepage zone, but it has a major impact on the height of the surface water within the seepage zone. Higher $\mu$ values correspond to lower surface water height, which in the limit $\mu\rightarrow\infty$ becomes negligible compared to the variation of the groundwater depth. Also, in this limit, the seepage zone size reaches a limiting value, $a_0=1-\frac{1}{\rho_0}$ (see \cref{sec:steady_state}). This limit is strongly supported by real-world data \edit{(typical value of $\mu$ for UK catchments is of the order $10^6$ based on the typical parameter values estimated in Part 1 of this study)}, and it will allow us to derive the formula for a typical hydrograph.
\end{enumerate}

\section{The formulation of an asymptotic model for intense rain}
\label{sec:asymptotic_analysis_high_rho0}

\noindent In the previous section, we presented numerical simulations of the full PDE system \eqref{eq:1Dmodel} and showed that under certain parameter choices, the resulting hydrographs could be approximately classified into two behaviours as shown in \cref{fig:early_late_time_behaviour}. In particular, when the system is initiated with an initial seepage zone, \emph{i.e.} $\rho_0 > 1$, then in response to an intense rainfall with $r > 1$, the river inflow rapidly increases over time. It is important to note that the duration of a standard intensive rainfall is much shorter compared to the typical timescale of the groundwater flow \edit{(\textit{i.e.} typical travel time along the hillslope)}, which is approximately 
\[
T_0=\frac{L_x}{K_sS_x} \approx \text{1000 days}.
\]
Our focus is to develop the short-time asymptotics to better understand this crucial response. Ultimately, we aim to derive an analytical solution for the river inflow, $Q(x=0,t)$. Based on the physical constraints, we are primarily interested in the following asymptotic limits:
\begin{equation} \label{eq:assumptions}
\begin{aligned}
    \text{small-time} &: \quad t \ll 1, \\
    \text{\edit{convection-dominated flow in the seepage zone}} &: \quad \mathrm{Pe} \gg 1, \\
    \text{intense rainfall} &: \quad \rho = r \rho_0 \gg 1.
\end{aligned}
\end{equation}

To begin, let us reformulate the governing system in terms of a boundary value problem. We assume that there exists a single contact line located at $x = a(t)$ where $H(a(t), t) = 1$. This configuration is illustrated in \cref{fig:1D_hillslope}. From \eqref{eq:dHdt_dimless}, the evolution of these surfaces is governed by:
\begin{subnumcases}
{\dt{H}=}
	\dx{}\left(\sigma\dx{H} + \mu (H-1)^k\right) + \rho & \quad $\text{for } x\in \left[0,a\left(t\right)\right]$,
    \label{eq:dHdt_saturated_zone} \\
        f(x)^{-1}\left[\dx{}\left(\sigma H\dx{H} + H\right) + \rho\right] & \quad $\text{for } x\in \left[a\left(t\right),1\right]$.
    \label{eq:dHdt_unsaturated_zone}
\end{subnumcases}
Here, we have introduced $\rho=\rho_0 r$. Thus, we have a set of two time-dependent equations for $H$ that are second-order in space, along with an additional contact-line position $a(t)$. Consequently, we require five boundary conditions in addition to the initial condition. Two boundary conditions are needed at $x = 0, 1$, and three matching conditions are required at the interface, $x = a(t)$. In total, these conditions are
\begin{subequations}\label{eq:bc_dimless}
    \begin{gather}
        \partial_x H(0, t) = 0, \qquad \partial_x H(1, t) = -\sigma^{-1},
        \tag{\theequation a-b} \\    
        H(a^-, t) = 1, \qquad H(a^+, t) = 1, \qquad \partial_x H(a^-, t) = \partial_x H(a^+, t),
        \tag{\theequation c-e}
    \end{gather}
\end{subequations}
where $a^\pm$ corresponds to the right/left limits as $x \to a$.

The first two boundary conditions are obtained from \eqref{eq:bc_river_v1} and \eqref{eq:no_flow_bc}. \edit{The next two boundary conditions arise from defining $a$ as the point where the groundwater table reaches the surface (\textit{i.e.} where $H=1$).} The last interface condition is a consequence of the continuity of flow given by~\eqref{eq:q_dimless}. Notice that a kinematic condition can be derived for the front position. Applying the chain rule to $H(a(t), t) = 1$, we have:
\begin{equation}
\pd{H}{t} + \pd{H}{x} \dd{a}{t} = 0 \qquad \text{at $x = a(t)$}.
\end{equation}
Following \eqref{eq:H_initial_condition}, we set the initial condition given by the steady state of eqs~\eqref{eq:dHdt_saturated_zone}-\eqref{eq:dHdt_unsaturated_zone} for $r=1$, which we denote as $H_0(x; \, \rho_0)$. Thus, $H_0$ satisfies:
\begin{equation}
    \label{eq:H_steady_state}
    0 = 
    \begin{dcases}
        \left(\sigma H_0H_0' + H_0\right)' + \rho_0 & \quad \text{for } x > a(t=0), \\
        \left(\sigma H_0' + \mu \left(H_0-1\right)^k\right)' + \rho_0 & \quad \text{for } x \leq a(t=0),
    \end{dcases}
\end{equation}
where primes $(')$ denote differentiation with respect to $x$. \edit{Here, $a(t=0)$ can be regarded as an eigenvalue and determined from this initial condition.}


\edit{In the next two sections, we will use this model to derive an asymptotic solution for the hydrograph $Q(t)$. Our approach involves three main steps:} 
\begin{enumerate}[label=(\roman*),leftmargin=*, align = left, labelsep=\parindent, topsep=3pt, itemsep=2pt,itemindent=0pt ]
\item First, in \cref{sec:steady_state}, we study the initial state $H(x, 0) = H_0(x; \, r = 1)$, which is assumed to be the steady-state response to the rain input $r=1$. This is a complicated coupled overland-groundwater problem, but we are able to develop analytical approximations in the limit of $\mu \to \infty$ or equivalently $\mathrm{Pe} \to \infty$.
\item Next, in \cref{sec:groundwater_rising}, we study the small-time response of the groundwater configuration and the propagation of the seepage zone relative to this initial steady state. At the time $t = 0$, the rainfall is set to $r > 1$, which causes the groundwater to rise and the seepage zone to shift. Analytical approximations can be developed for the case of $\Pe \to \infty$ and for large rainfalls, $r \to \infty$. 
\item Finally, in \cref{sec:method_of_characteristics}, we develop an analytical approach for predicting the evolution of the overland flow, which leverages our analysis of the seepage zone propagation obtained in (ii). This turns out to be a wave propagation study using the method of characteristics.
\end{enumerate}

\section{Asymptotic analysis of the initial condition, \texorpdfstring{$H_0$}{}, with \texorpdfstring{$\mathrm{Pe} \to \infty$}{}}
\label{sec:steady_state}

\noindent In our model, we assume that the system begins at the configuration that corresponds to the particular steady-state solution forced by the 'typical' rainfall, $\rho_0$.

By integrating~\eqref{eq:H_steady_state} and applying the upstream boundary condition $q(1)=0$, we obtain:
\begin{subnumcases}{\rho_0(1-x)=}
    H_0 + \sigma H_0 H_0' & for $x > a$,
    \label{eq:Hg_steady_state_integrated} \\
    1 + \sigma H_0' + \mu(H_0-1)^k & for $x \leq a$.
    \label{eq:Hs_steady_state_integrated}
\end{subnumcases}

The fact that the limit $\mu \to \infty$ involves the P\'{e}clet number, defined as $\Pe = \mu^{1/k}/\sigma$ via \eqref{eq:peclet}, is not entirely obvious. Note that as $\mu \to \infty$, the dominant balance in the overland equation for $x \leq a$ indicates that $H_0 \sim 1$ in this limit. We re-scale $x = aX$ and $H_0 = \mu^{-1/k} g(X)$, obtaining, for $X \in [0, 1]$,
\begin{gather}
        \frac{\Pe ^{-1}}{a}\frac{\partial g}{\partial X}+g^k=\rho_0(1-aX) - 1, \label{eq:problem_1}\\
        g'(0) = 0 \quad \text{and} \quad g(1) = 0.
\end{gather}

\edit{In the limit $\Pe\to\infty$, we note that naively, the diffusion term in \eqref{eq:problem_1} tends to zero. Then, since $g(1) = 0$, we can approximate $\rho(1 - a) \sim 1$, which gives the front position as $a\sim 1-1/\rho_0$. However, note that in this limit, the leading (outer) solution is given by $g \sim [\rho_0 (1 - aX) - 1]^{1/k}$, and hence exhibits an infinite gradient as $X \to 1$. Consequently, it is not obvious that the diffusion term can be neglected \emph{a priori} as $\Pe\to\infty$. The gradient of the solution exhibits a boundary layer and thus requires a matched asymptotics approach.}

In \cref{app:steady_state}, we show that the contact line, $x = a$, and the gradient at the front can be expanded into an asymptotic expansion. In terms of the original $H_0$, this is
\begin{equation}
    \label{eq:adh}
    a\sim a_0+a_1\Pe ^{-\beta} \quad \text{and} \qquad H_0'(a) \sim \left[-\frac{\rho_0 a_1}{\sigma}\right] \Pe ^{-\beta},
\end{equation}
where $\beta = k/(2k-1)$ and the leading-order contact position is indeed
\begin{equation}
    \label{eq:a0}
     a_0 = 1-\frac{1}{\rho_0}.
\end{equation}
Notice that increasing the rainfall rate, $\rho_0 \to \infty$, sends $a \to 1$, and overland water saturates the entire hillslope. In contrast, the limit $\rho_0 \to 1^+$ reduces the seepage zone size to zero, as anticipated in \cref{sec:typical_dynamics}. The correction factor of $a_1$ in \eqref{eq:adh} can be calculated as an eigenvalue via the solution of a boundary-value problem [cf. eqn \eqref{eq:bvp_for_a1}]. Finally, notice that as $\Pe ^{-1} \to 0$, the gradient at the transition between overland and groundwater flows, $H_0'(a) \to 0$.

As we shall show in \cref{sec:groundwater_rising}, in order to find the speed of the seepage zone growth, we need to find the initial depth of the groundwater outside the seepage zone first. We can find this initial depth by solving~\eqref{eq:Hg_steady_state_integrated}, which can be rearranged to
\begin{equation}
    \label{eq:H0_ode}
    \sigma \dd{H_0}{x} = \frac{\rho_0(1-x)}{H_0(x)} - 1\quad \text{for } x\in[a_0,1],
\end{equation}
with a boundary condition $H_0(a_0)=1$. This first-order nonlinear ODE does not have an explicit analytical solution; it can either be solved numerically, or we can investigate its shape in different limits. 

\subsection{Analytical solution as \texorpdfstring{$\sigma \to 0$}{}}

\begin{figure}
    \centering
    \includegraphics{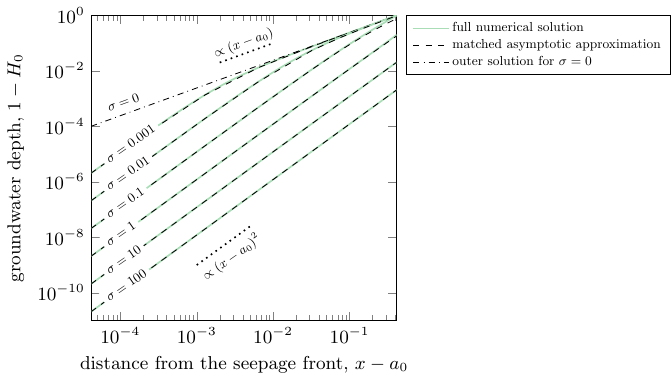}
    \caption{Comparison of groundwater depth given by~\eqref{eq:H0_ode} (full numerical solution) to the matched asymptotic approximation given by~\eqref{eq:H0_matched_solution}.}
    \label{fig:groundwater_models}
\end{figure}

\noindent One quite useful limit is to consider $\sigma \rightarrow 0$, corresponding to the infinitely thin porous layer limit. In \cref{app:H0_derivation}, we derive the outer asymptotic expansion for $H_0$ in terms of $\sigma$, \eqref{eq:outer_expansion_H0}, its inner expansion around $x=0$ \eqref{eq:inner_expansion_H0}, and finally match them to form the following composite approximation for $H_0(x)$:
\begin{equation}
    \label{eq:H0_matched_solution}
    H_0(x)=\rho_0\left(1-x+\sigma-\sigma e^{-\frac{x-a_0}{\sigma}}\right), \qquad \text{for $x \in [a_0, 1]$.}
\end{equation}
As shown in \cref{fig:groundwater_models}, this asymptotic solution provides a good approximation of the groundwater shape both for small $\sigma$ values and, surprisingly, also for large $\sigma$ values. In the latter case, $H_0$ becomes a \editt{quadratic function,
\begin{equation}
    1-H_0\sim \frac{\rho_0}{2\sigma}\left(x-a_0\right)^2,
\end{equation}}
\noindent which is also a limiting behaviour of our matched asymptotic solution~\eqref{eq:H0_matched_solution} \editt{as $x\to a_0$}.

\section{Short-time asymptotics}
\label{sec:short_time_asymptotics}

\noindent This section relates to the asymptotic limits of $t \to 0$, $\mu \to \infty$, and $\rho = \rho_0 r \to \infty$ in \eqref{eq:assumptions}. 

\subsection{Groundwater rise and propagation of the seepage zone}
\label{sec:groundwater_rising}

\noindent Having derived certain analytical properties of the steady-state configuration, $H_0(x; \, \rho_0)$ (used as an initial condition), we can now study the short-time behaviour of the system as the rain input is set to $\rho$. As argued at the start of \cref{sec:asymptotic_analysis_high_rho0}, this is a good approximation, \editt{when the rainfall duration is much shorter than the characteristic time of the groundwater transfer to the channel}.

\begin{figure}
  \centering
  \includegraphics{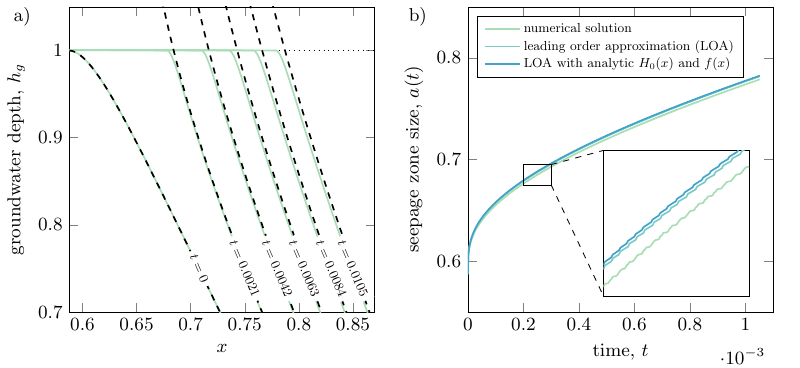}
  \captionof{figure}{(a) Comparison of numerical solution of the groundwater shape (solid lines) with the outer solution developed in \cref{app:groundwater_outer_derivation} (dashed lines) at different times $t$. The corresponding size of the seepage zone is presented in (b). A small region is magnified to highlight differences between the presented approximations. The lines are not smooth due to the $h(x)$ interpolation error.}
  \label{fig:a_vs_t}
\end{figure}

As shown in \cref{app:groundwater_outer_derivation}, the outer solution outside the seepage zone~\eqref{eq:dHdt_unsaturated_zone} can be expanded into a regular series expansion in powers of time $t$: 
\begin{equation} \label{eq:Hexpand_inittime}
    H_\text{outer}(x,t) \sim H_0(x; \, \rho_0) + \left[\frac{\rho - \rho_0}{f(x)}\right] t + \Oh(t^2).
\end{equation}
The above approximation assumes that $x - a(t) = \Oh(1)$. The approximation \eqref{eq:Hexpand_inittime} thus indicates that the groundwater rises in a fashion proportional to time and the difference between current and prior rain input; it correctly describes the shape of the groundwater except for a thin boundary layer at $x=a(t)$ of thickness of $\Oh(\sqrt{t/(\rho - \rho_0)})$ (see \cref{fig:a_vs_t}a). Therefore, for intense rainfall, $\rho\gg 1$, we can neglect the effect of this boundary layer.

We may use the outer groundwater approximation, \eqref{eq:Hexpand_inittime}, in order to predict the motion of the contact line, $x = a(t)$. Setting $H_\text{outer} = 1$ gives, in implicit form:
\begin{equation}
    \label{eq:front_equation}
    t \sim \frac{f(x=a(t))}{\rho - \rho_0} \Big(1 - H_0\big(x=a(t)\big)\Big) \equiv \mathcal{T}(x=a(t)).
\end{equation}
In order to calculate the above, we must solve two first-order ODEs: \eqref{eq:H0_ode} for the height, $H_0(x)$, and \eqref{eq:Richards_steady_state} for the head, $h_g(\hat{z})$, itself used in the calculation of $f(x)$. Alternatively, one can use the analytical approximations for $H_0(x)$ given by~\eqref{eq:H0_matched_solution}, and $f(x)$ given by~\eqref{eq:f_approx_1} or \eqref{eq:f_approx_2}. Fig.~\ref{fig:a_vs_t}b compares these approximations with the location of the saturation front computed from a full numerical solution of the 1D model. As we observe \edit{based on the difference between the full numerical solution and leading-order approximation (LOA)}, neglecting the boundary layer around $a(t)$ introduces a small error when estimating the seepage zone size. Replacing the ODEs with analytical approximations for $f(x)$ and $H_0(x)$ in \eqref{eq:front_equation} also introduces an error, but it is significantly smaller.

\subsection{Evolution of the overland flow}
\label{sec:method_of_characteristics}

\noindent Now, knowing how the seepage zone propagates, we can develop a time-dependent solution for the overland flow. Our goal is to extract how the overland flow into the river, $Q_s(x=0, t)$, evolves in time, taking into account the effects of increased rainfall and the seepage zone growth.

\subsubsection{Problem reduction under $\Pe\rightarrow\infty$ limit}

\noindent The equation for overland flow is given by~\eqref{eq:dHdt_saturated_zone} with the initial condition satisfying steady state~\eqref{eq:Hs_steady_state_integrated}. We re-scale according to:
\begin{equation}
\eta = \mu^{1/k}(H-1) \quad \text{and} \quad T = \mu^{1/k}t.
\end{equation}
Here, $\eta = \eta(x, T)$ is the re-scaled surface water height $h_s=H-1$. Then \eqref{eq:dHdt_saturated_zone} can be written as:
\begin{equation}
    \label{eq:dhdt_overland}
    \dT{\eta} - k \eta^{k-1}\dx{\eta} - \Pe ^{-1}\ddx{\eta}= \rho,
\end{equation}
and equation~\eqref{eq:Hs_steady_state_integrated}, which provides the initial condition, $H_0$, is:
\begin{equation}
    \label{eq:hs_initial}
    1 + \eta^k - \Pe^{-1} \dd{\eta}{x} = \rho_0 (1 - x).
\end{equation}
where, as before, $\Pe ^{-1}=\sigma/\mu^{1/k}$. \edit{Following (\ref{eq:bc_dimless}a) and (\ref{eq:bc_dimless}c) the boundaries conditions are:
\begin{subequations}\label{eq:bc_overland}
    \begin{gather}
        \partial_x \eta(0, t) = 0, \qquad \eta(a(t), t) = 0.
        \tag{\theequation a-b}
    \end{gather}
\end{subequations}}

Note that the characteristic time \editt{it takes the overland flow to reach the channel} ($\mu^{-1/k}T_0\approx\text{0.1 day}$) is much shorter than the characteristic time describing the groundwater \editt{flow} ($T_0\approx\text{1000 days}$), and has a similar order of magnitude as a typical rainfall duration. As a result, a short-time approximation is not satisfactory to describe flow variation during a single rainfall event.

The solution for general times can be obtained by considering the $\Pe \rightarrow \infty$ limit, similarly as we did when analysing the steady state. This limit allows us to neglect the diffusion term everywhere except for a negligibly thin boundary layer around $x=a$.

In the limit $\Pe \rightarrow \infty$, we expand $\eta = \eta_0 + \Pe^{-1} \eta_1 + \ldots$, and equation~\eqref{eq:dhdt_overland} becomes a first-order hyperbolic PDE:
\begin{subequations} \label{eq:overland_sys_char}
\begin{equation}
    \dT{\eta_0} - \Bigl[k \eta_0^{k-1}\Bigr]\dx{\eta_0}= \rho, \qquad x \geq 0. \label{eq:dhdt_overland_limit} 
\end{equation}
For $0 \leq x \leq a(t)$, there is an initial condition given by 
\begin{equation}
\eta_0(x, 0) = \rho_0^{1/k}\left(a_0-x\right)^{1/k}, \label{eq:h0}
\end{equation}
\end{subequations}
where we have used the fact shown in \cref{app:steady_state} that $a_0=1-1/\rho_0$ [cf.  \eqref{eq:a0}]. The above initial condition is defined along the entire initial seepage zone, $x\in[0,a_0]$. \edit{Note that neglecting the diffusion term results in a kinematic wave equation, for which the downstream boundary condition (\ref{eq:bc_overland}a) is no longer required.}

\subsubsection{Implicit solution using methods of characteristics}

\begin{figure}
    \centering
    \includegraphics{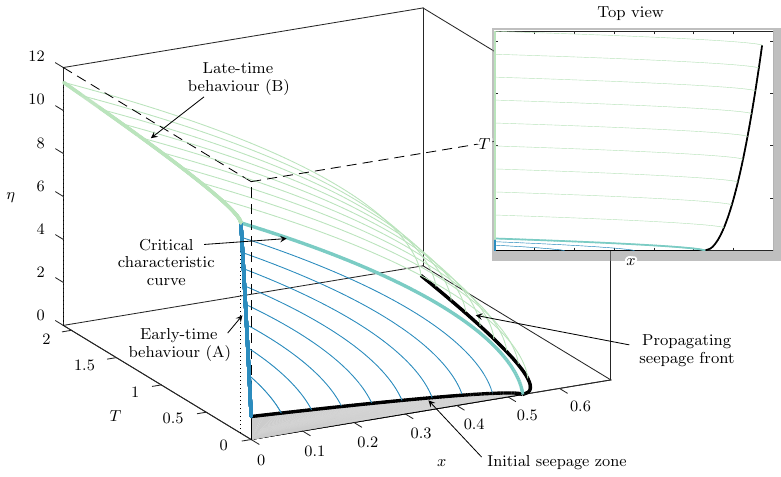}
    \caption{Characteristic curves given by~\eqref{eq:characteristic_curves} for parameters listed in~\cref{tab:typical_parameters}. \edit{Dark blue} lines represent curves originating from the initial seepage zone, and \edit{light green} lines represent curves originating from the propagating front of the seepage zone.}
    \label{fig:characteristics_diagram}
\end{figure}

\noindent The system \eqref{eq:overland_sys_char} can be solved using the method of characteristics (Lagrange-Charpit equations). The solution is given by characteristic curves $(T, x, \eta_0)$, now parameterised by $(s, \tau)$, where $\tau$ is the characteristic curve parameter, and $s$ parameterises the initial data. The characteristic equations are:
\begin{subequations} \label{eq:charpit_sys}
\begin{equation}
    \label{eq:Lagrange_Charpit}
    \dd{T}{\tau}=1, \qquad
    \dd{x}{\tau}=-k\eta_0^{k-1}, \qquad
    \dd{\eta_0}{\tau}=\rho.
\end{equation}
The initial conditions are specified along $\tau = 0$ according to two types of characteristics. One set of characteristics emerges from $T = 0$, at the location of the initial water shape, $H_0(x)$, valid for $x\in [0, a(t)]$. Another set of characteristics emerges from the propagating front, $x = a(t)$, representing the groundwater reaching the surface and hence initiating surface flow. 

Parameterising the initial data by $x = s$, we have:
\begin{gather}
(T(s, 0), x(s, 0), \eta_0(s, 0)) = 
\begin{cases}
\Bigl(0, s, H_0(s)\Bigr), & s \in [0, a(t)], \\
\Bigl(\mu^{1/k}\mathcal{T}(s), s, 0\Bigr), & s \in [a(t), \infty).
\end{cases}
\label{eq:characteristics_IC}
\end{gather}
\end{subequations}
The first condition will use the initial surface height, $H_0(s) = \rho_0^{1/k}\left(a_0-s\right)^{1/k}$ given by~\eqref{eq:h0}. The second condition is essentially specified along the moving front, $(T, x, \eta_0) = (T, a(t), 1)$, but we have written it in terms of the $s$-independent variable, and the rescaled function $\mathcal{T}$ in \eqref{eq:front_equation}. In summary, the characteristic solution can be obtained via direct integration of \eqref{eq:charpit_sys}, giving:
\begin{equation}
    \label{eq:characteristic_curves}
    \begin{split}
        T(s,\tau)&=T(s,0)+\tau, \\        x(s,\tau)&=x(s,0)-\rho^{-1}\big[\eta_0(s,0)+\rho\tau\big]^k+\rho^{-1}\big[\eta_0(s,0)\big]^k, \\
        \eta_0(s,\tau)&=\eta_0(s,0)+\rho\tau.
    \end{split}
\end{equation}
We show an example of the characteristics and characteristic projections in \cref{fig:characteristics_diagram}.

Once the solution is determined, a key quantity of interest is the surface water height at $x = 0$, as it determines the overland flow reaching the river. We denote this critical point along the characteristics as $(T, \eta_0) = (T^*, \eta^*)$. By setting $x(s,\tau)=0$ in the characteristic equations \eqref{eq:characteristic_curves} and eliminating $\tau$ from the second equation, we obtain:
\begin{equation}
    \label{eq:hs_at_river}
    \begin{pmatrix}
    T^* \\ 
    \eta^*
    \end{pmatrix}
    =
    \begin{pmatrix}
    T(s,0)+\frac{1}{\rho}\big(\eta^*-\eta_0(s,0)\big) \\
    \left(\rho x(s,0)+\big(\eta_0(s,0)\big)^k\right)^{1/k}
    \end{pmatrix}.
\end{equation}
We need to consider two cases separately: characteristics starting from either the initial seepage zone, and characteristics emerging from the propagating seepage front. Each case is given by the different initial conditions, as specified in \eqref{eq:characteristics_IC}.

In the first case, by substituting the first initial condition from~\eqref{eq:characteristics_IC} into~\eqref{eq:hs_at_river}, we obtain:
\begin{equation}\label{eq:hs_1}
\begin{pmatrix}
    T^* \\ 
    \eta^*
    \end{pmatrix}
    =
\begin{pmatrix}
    \frac{1}{\rho}\left(\eta^*-\rho_0^{1/k}\left(a-s\right)^{1/k}\right) \\
    \left(\rho s+\rho_0\left(a-s\right)\right)^{1/k}
\end{pmatrix}.
\end{equation}
By finding $s$ from (\ref{eq:hs_1}a) and substituting to (\ref{eq:hs_1}b) we can express $T^*$ as a function of $\eta^*$:
\begin{equation}
    \label{eq:t_start_case1}
    T^*(\eta^*)=\frac{1}{\rho}\left[\eta^*-\left(\frac{\rho_0}{\rho-\rho_0}\right)^{1/k}\left(\rho a_0 - \left(\eta^*\right)^k\right)^{1/k}\right],
\end{equation}
This equation is satisfied for $\eta^*\in[(\rho_0 a_0)^{1/k}, (\rho a_0)^{1/k}]$. The lower limit corresponds to the the initial height, and the upper limit corresponds to the height reached by the characteristic curve starting at $x_0=a_0$. At the upper limit, the characteristic curve reaches the river ($x=0$) at what we refer to as the \emph{critical time}:
\begin{equation}
    \label{eq:tsat}
    \Tsat=\frac{1}{\rho}(\rho a_0)^{1/k}.
\end{equation}
This critical saturation event is associated with the critical characteristic curve highlighted in \cref{fig:characteristics_diagram}.

For those characteristics starting from the propagating seepage front, $x = a(t)$, we substitute the second initial condition from~\eqref{eq:characteristics_IC} into~\eqref{eq:hs_at_river}:
\begin{equation}
\begin{pmatrix}
    T^* \\ 
    \eta^*
    \end{pmatrix}
    =
    \begin{pmatrix}
    \mu^{1/k}\mathcal{T}(s)+\frac{1}{\rho}\eta^* \\
    \left(\rho s\right)^{1/k}
    \end{pmatrix}.
\end{equation}
By eliminating $s$, we can express $T^*$ as a function of $\eta^*$:
\begin{equation}
    \label{eq:t_start_case2}
    T^*(\eta^*) =\mu^{1/k} \mathcal{T}\left(\frac{1}{\rho}\left(\eta^*\right)^k\right)+\frac{1}{\rho}\eta^*.
\end{equation}

By combining equations~\eqref{eq:t_start_case1} and~\eqref{eq:t_start_case2}, we can find the height of the surface water, at the river, $x = 0$, value for all times $T \geq 0$. This is done by solving the implicit equation:
\begin{equation}
    \label{eq:Q_implicit_1}
    T^*(\eta^*) = 
    \begin{cases}
        \frac{\eta^*}{\rho}-\frac{1}{\rho}\left(\frac{\rho_0}{\rho-\rho_0}\right)^{1/k}\left(\rho a_0 - \left(\eta^*\right)^k\right)^{1/k}, & \text{for } \eta^* \leq (\rho a_0)^{1/k}, \\
        \frac{\eta^*}{\rho}+\mu^{1/k} \mathcal{T}\left(\frac{1}{\rho}\left(\eta^*\right)^k\right), & \text{for } \eta^* > (\rho a_0)^{1/k}.
    \end{cases}
\end{equation}
Alternatively, we can express the height of the surface water $\eta^*$ in terms of the overland component of river inflow, which is represented by the last term in~\eqref{eq:q_dimless}, $Q_s^*=\left(\eta^*\right)^{\edit{k}}$. This leads to the equation:
\begin{equation}
    \label{eq:Q_implicit}
    t^*(Q_s^*) = 
    \begin{cases}
        \frac{1}{\rho}\left(Q_s^*\right)^{1/k}-\frac{1}{\rho}\left(\frac{\rho_0}{\rho-\rho_0}\right)^{1/k}\left(\rho a_0 - Q_s^*\right)^{1/k}, & \text{for } Q_s^* \leq \rho a_0, \\
        \frac{1}{\rho}\left(Q_s^*\right)^{1/k}+\mu^{1/k} \mathcal{T}\left(\frac{Q_s^*}{\rho}\right), & \text{for } Q_s^* > \rho a_0.
    \end{cases}
\end{equation}
Equation \eqref{eq:Q_implicit} represents one of the major results of this work, since it provides an implicit expression for the shape of the hydrograph $Q_s^*(t^*)$.

\subsubsection{Approximating the hydrograph in an explicit form}
\label{sec:explicit}

\noindent We can obtain an approximated explicit form for the $Q_s^*(t)$ function for $Q_s^* > \rho a_0$. In the limit $\mu \to \infty$ (equivalent to $\Pe \to \infty$), we may expand around the value of $Q_s^*$ at $\tsat$ in \eqref{eq:tsat}, and write:
\begin{subequations} \label{eq:qs_explicit_group}
\begin{equation}
    \label{eq:qs_explicit}
    Q_s^*(t^*) \sim \rho a\left(\mu^{-1/k}\left(t^*-\tsat\right)\right)\quad\text{for } t^* \geq \tsat,
\end{equation}
where we have used $a(t)=\mathcal{T}^{-1}(t)$ from \eqref{eq:front_equation} to describe the propagation of the wetting front in time. Following approximation~\eqref{eq:H0_matched_solution} and~\eqref{eq:f_approx_2}, it can be written explicitly as:
\begin{equation}
    \label{eq:qs_explicit_a}
    a(t)=\underbrace{a_0}_{\text{term 1}}+\underbrace{s(t)}_{\text{term 2}}+\underbrace{\sigma\left[1+W_0\left(-\mathrm{e}^{-1-s(t)/\sigma}\right)\right]}_{\text{term 3}},
\end{equation}
where
\begin{equation}
    \label{eq:qs_explicit_b}
    s(t)=A t^{\frac{1}{n + 1}} \quad\text{with}\quad A = \frac{1}{\rho_0} \left[\frac{n + 1}{m} \frac{\rho-\rho_0}{\theta_s-\theta_r}\left(1-\frac{r_0}{K_s}\right)^{-n} \alpha^{-n}\right]^{\frac{1}{n + 1}}.
\end{equation}
\end{subequations}
Here, $W_0(\cdot)$ is the Lambert W function, \edit{and $\alpha$, $\theta_s$, $\theta_r$, $n$, and $m$ are soil properties used in the Mualem Van-Genuchten model (see \cref{sec:mean_drainable_porosity})}.

The first term in~\eqref{eq:qs_explicit_a} corresponds to the flow over the initial seepage zone. The second and third terms represents the growth of flow after reaching the critical point. The third term, in the case of $\sigma\ll 1$, quickly grows from $0$ asymptotically reaching $\sigma$ as $t\rightarrow\infty$, while the second term is responsible for further growth of the river flow. Therefore, for thin hillslopes ($\sigma\ll 1$), the growth of river flow after passing the critical point scales proportionally to the $\rho A$ factor.

\editt{In the next section, we summarise all approximations derived in this section, and validate them by comparing to each other and full numerical solutions of 1D and 2D benchmark models.}

\section{A numerical comparison between different approximations}
\label{sec:numerical_comparison}

\subsection{Numerical setup}

\noindent In Part 2~\citep{paper2}, we performed a numerical verification of the assumption of reducing the three-dimensional benchmark model to a two-dimensional model. We also conducted a detailed sensitivity analysis, highlighting the dependencies of model parameters on the resultant peak flows. Here, we continue this analysis by comparing the hydrographs and peak flows between the five different approximations derived and discussed in this paper. The approximations are summarised in \cref{tab:approximations} in order from the most complex to the simplest. 

\begin{table}
    \centering \footnotesize
    \def\arraystretch{1.5}%
    \begin{tabular}{rp{7cm}}
        \textsc{Approximation} & \textsc{Equations/notes}\\
        2D surface-subsurface model & Richards/Saint Venant equations along a 2D hillslope, as discussed in Part 2 of this paper. \\
        1D surface-subsurface model & The 1D Boussinesq system~\eqref{eq:1Dmodel}, which assumes a thin porous layer limit, $L_z \ll L_x$. \\
        Characteristics (numerically implicit) & Solution given by~\eqref{eq:Q_implicit}, where the $H_0(x)$ and $f(x)$ functions are found numerically using 1D ODEs \eqref{eq:H0_ode} and~\eqref{eq:Richards_steady_state}. Assumes scenario $\rho_0>1$, early time $t\ll 1$, intense rainfall $\rho\gg 1$, and $\Pe\gg 1$. \\
        Characteristics (analytically implicit) & Solution given by~\eqref{eq:Q_implicit}, where the functions $H_0(x)$ and $f(x)$ are approximated as~\eqref{eq:H0_matched_solution} and \eqref{eq:f_approx_1}. In addition, it assumes that $\sigma\gg 1$, $H_0\ll 1$. \\
        Characteristics (analytically explicit) & Solution given by \eqref{eq:qs_explicit_group}, $\tsat<t\ll 1$ (in addition to the assumptions listed before).\\
        Critical flow & Flow estimated as $\qsat=\rho a_0 + \rho_0 (1-a_0)$, equal to the river inflow reached at $t=\tsat$ (further discussion in \cref{sec:hydrograph_features}).
    \end{tabular}
    \caption{Summary of the approximations developed in this work.}
    \label{tab:approximations}
\end{table}

Similar to the methodology presented in Parts 1 and 2, we assess the performance of the above models in two ways. Firstly, we compare the hydrograph obtained using each model for standard values of parameters characterising UK catchments, as listed in \cref{tab:simulation_settings}. Secondly, we run a sensitivity analysis by varying seven model parameters, one at a time, while keeping the others at their default values, and measuring the peak flow in the river after an intensive rainfall. In both numerical experiments, we consider a uniform rainfall over a duration of 24 hours.

\begin{table} \centering
    \begin{tabular}{cccc|}
        \textsc{parameter} & \textsc{default value} & \textsc{parameter range} \\
        $K_s\;[\mathrm{ms^{-1}}]$ & $1\cdot 10^{-5}$ &  $10^{-6}-10^{-4}$ & \phantom{xx} \\
        $L_x\;[\mathrm{m}]$ & $6.16\cdot 10^2$ & $10^2-10^3$ &\\
        $L_z\;[\mathrm{m}]$ & $6.84\cdot 10^2$ & $10^1-10^3$ &\\
        $S_x\;[-]$ & $7.5\cdot 10^{-2}$ & $10^{-2}-10^{-1}$ &\\
        $r\;[\mathrm{ms^{-1}}]$ & $2.36\cdot 10^{-7}$ & $3\cdot 10^{-8}-3\cdot 10^{-6}$ &\\
        $r_0\;[\mathrm{ms^{-1}}]$ & $2.95\cdot 10^{-8}$ & $10^{-9}-10^{-7}$ &\\
        $n_s\;[\mathrm{ms^{-1/3}}]$ & $5.1\cdot 10^{-2}$ & $10^{-2}-10^{-1}$ &
    \end{tabular}
    \begin{tabular}{cc}
        \textsc{parameter} & \textsc{value} \\
        $L_y\;[\mathrm{m}]$ & $18000$ \\
        $w\;[\mathrm{m}]$ & $5$ \\
        $h_\mathrm{out}\;[\mathrm{m}]$ & $0.3$ \\
        $\theta_s\;[-]$ & $0.488$ \\
        $\theta_r\;[-]$ & $0$ \\
        $n\;[-]$ & $1.19$ 
    \end{tabular}
    \caption{Default values and ranges of parameters used to perform the sensitivity analysis. The table on the right presents parameters not varied during the sensitivity analysis.}
    \label{tab:simulation_settings}
\end{table}

\subsection{Comparing the hydrographs}

\begin{figure}
    \centering
    \includegraphics{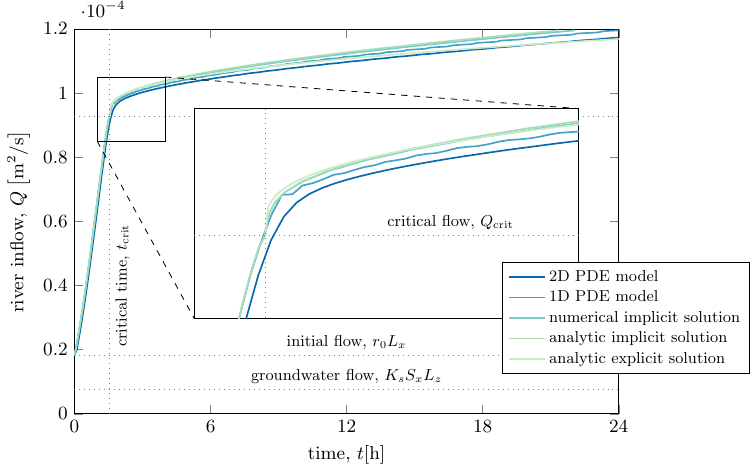}
    \caption{Hydrograph computed using approximations listed in \cref{tab:approximations} for default values of parameters given in \cref{tab:simulation_settings}. The graph area around the critical point is magnified. \edit{Numerical instability are observed for the 1D model, caused by the finite discretisation of space, which does not allow capturing the exact location of the seepage, and by instabilities related to the governing equation for the seepage zone \eqref{eq:dHdt_dimless}, characterised by a very small diffusive term.}}
    \label{fig:hydrograph_approximations}
\end{figure}

\noindent A comparison of the hydrographs obtained under the different approximations is presented in \cref{fig:hydrograph_approximations}. Firstly, we note that the 1D models formulated in this paper produce similar hydrographs to the 2D model from the previous part of our work. However, the 1D models slightly underestimate the flow, and the solutions are not smooth around the critical point separating early-time and last-time growth.

Secondly, all approximated solutions of the 1D model produce consistent results for $t\leq\tsat$ (except for the explicit solution, which is valid only for $t>\tsat$). The results are also similar for $t>\tsat$, but inaccuracies related to different approximations start to become noticeable. For example, the implicit solution closely follows the 1D model solution, but with the flow slightly shifted towards the higher values. This deviation is caused by neglecting the boundary layer characterising the groundwater shape around the critical point. As discussed in \cref{app:groundwater_outer_derivation}, this inaccuracy decreases as $\rho$ increases. Replacing numerical solutions for $f(x)$ and $H_0(x)$ with their analytical approximations seems to have a negligible impact on the model for typical sizes of catchment parameters.

Similarly, using approximation~\eqref{eq:qs_explicit_group} for the explicit solution leads to the underestimation of the groundwater rise rate, which slows down the growth of flow for higher $t$ values. Additionally, the flow around $t=\tsat$ is slightly overestimated as a result of neglecting the variation of the $\left(Q_s^*\right)^{1/k}$ term appearing in the implicit solution~\eqref{eq:Q_implicit}. Despite these small inaccuracies, the explicit solution still seems to produce excellent qualitative and quantitative agreement. Moreover, due to its simple form, the explicit form allows us to directly understand the impact of various catchment properties on the expected peak flows.

\subsection{Sensitivity analysis}

\begin{figure}
    \centering
    \includegraphics[width=\linewidth]{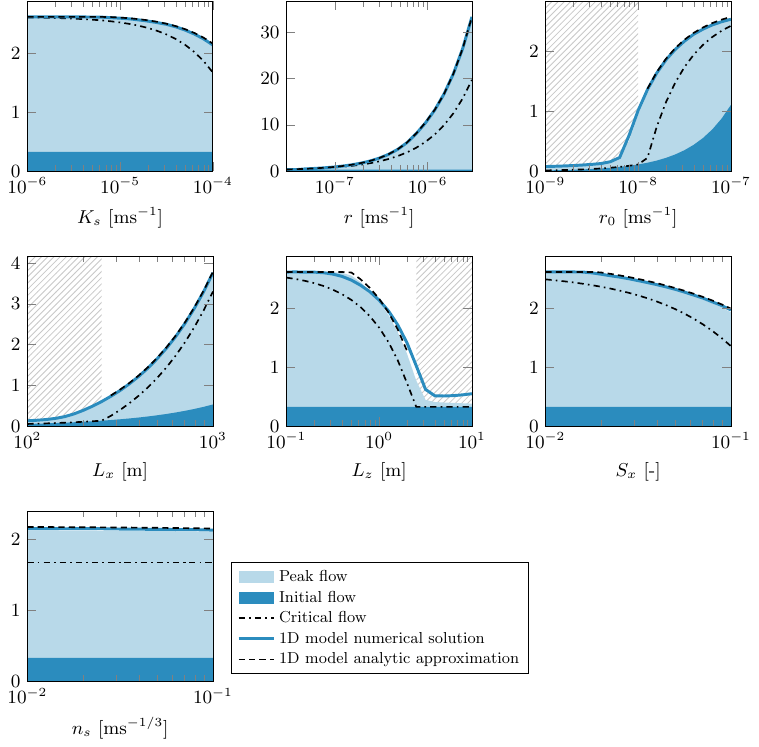}
    \caption{Sensitivity analysis results showing the dependence between the peak flow reached after 24 hours of intensive rainfall and seven different model parameters. The predictions conducted using four models of varying complexity are presented. The dashed region represents the parameter range, for which there is no initial seepage zone ($\rho_0<1$).}
    \label{fig:sensitivity_summary_scenario_B}
\end{figure}

\noindent We chose seven physical parameters for the sensitivity analysis: catchment width $L_x$, aquifer depth $L_z$, elevation gradient along the hillslope $S_x$, hydraulic conductivity $K_s$, precipitations rates $r$ and $r_0$, and Manning's constant $n_s$. We varied each parameter within the range of its typical values presented in \cref{tab:simulation_settings}, while keeping the other parameters constant. \edit{In each case, we simulated the model's response to a 24-hour-long rainfall event (as shown in \cref{fig:hydrograph_approximations}), and then measured the peak river inflow $Q(x=0,t)$ reached at the end of this period.} The results of the sensitivity analysis are presented in \cref{fig:sensitivity_summary_scenario_B}. 

We note that the dimensionless parameter determining the existence of the initial seepage zone is given by $\rho_0 = r L_x/(K_s L_z S_x)$. Therefore, as the dimensional parameters are varied, the initial condition may not involve an initial seepage zone if $\rho_0 < 1$. The parameter ranges for which that happens are marked by a dashed region in \cref{fig:sensitivity_summary_scenario_B}. We observe that as long as there is an initial seepage zone, the analytical approximations of the hydrograph are largely accurate over the range of tested parameters. 

As expected, the (total) peak flows reached are higher than the maximum levels set by the critical saturation flow, given by \eqref{eq:qsat_dimensional}, since the latter only describes the flow reached during the early-time phase. Nevertheless, for most parameter values, the critical flow curve provides a good approximation of the peak flows. In Part 2 of our work, based purely on the 3D and 2D simulations, we arrived at the same conclusion. 

The cases where the critical flow value highly underestimates the peak flow are around $\rho_0=1$. \edit{In these cases, either the seepage zone does not initially exist but the soil is almost fully saturated near the river ($\rho_0$ slightly lower than 1), or it exists but is very small ($\rho_0$ slightly larger). In both cases, rainfall causes the seepage zone to grow significantly relative to its initial size; however, this growth is not captured by the time-independent estimate \eqref{eq:qsat_dimensional}.} 

\section{Summary of the key hydrograph features}
\label{sec:hydrograph_features}

\noindent In \cref{fig:hydrograph_approximations} in the previous section, we showed a typical shape of a hydrograph given by the 1D model. Here, we summarise the main features of the hydrograph and its importance in benchmarking.

As discussed before, two different phases are visible: (Phase 1) an early-time fast rise caused by water accumulating over an initial seepage zone, and (Phase 2) a late-time slow rise caused by a growing seepage zone. The analytical approximation presented in \cref{sec:explicit} shows that \edit{the growth of the overland flow in the second phase can be approximated as $Q_s(t)\approx \rho a(t)$, \emph{i.e.} it corresponds to the total precipitation rate over the seepage zone $a(t)$ slowly growing in time $t$. Together with the groundwater flow $Q_g=1$, they give a total river inflow $Q(t)=1+\rho a(t)$, which in dimensional units is:
\begin{equation}
    \label{eq:Q_dimensional}
    Q(t) = K_s S_x L_z + r L_x a(t) = r_0 L_x \Big(1 - a_0\Big) + r L_x a(t)
\end{equation}
where $a_0=1-\frac{K_sS_xL_z}{rL_x}$ is the size of the initial seepage zone.}

Following~\cref{sec:method_of_characteristics}, the transition from the first to the second phase corresponds to the moment when the characteristic curve starting from the further-most point of the initial seepage zone reaches the river ($x=0$). This observation allowed us to estimate the dimensionless critical flow, which in dimensional units correspond to~\eqref{eq:Q_dimensional} with $a(t)=a_0$:
\begin{equation}
    \label{eq:qsat_dimensional}
    \qsat = \underbrace{K_s S_x L_z}_{\text{groundwater flow}} + \underbrace{r L_x \left(1 - \frac{K_s S_x L_z}{r_0 L_x}\right)}_{\text{overland flow}},
\end{equation}
It is reached at the critical time $\tsat$ given by~\eqref{eq:tsat}, which in dimensional units is:
\begin{equation}
    \tsat = \frac{L_z}{r}\left[\frac{S_x^{1/2}K_s n}{L_z^{k-1}} \left(\frac{L_x r}{K_s S_x L_z} - \frac{r}{r_0}\right)\right]^{1/k}.
\end{equation}

Following the above event, further growth~\eqref{eq:Q_dimensional} is slow, which is a result of the difference of $\mu^{1/k}\approx 10^4$ factor between the characteristic timescale of overland flow (responsible for Phase 1) and groundwater flow (responsible for Phase 2). \edit{Therefore, $\qsat$ may be a good approximation of the flow even long after the critical time.}

We highlight a few additional features \editt{of our 1D benchmark model}:
\begin{enumerate}[label={(\roman*)},leftmargin=*, align = left, labelsep=\parindent, topsep=3pt, itemsep=2pt,itemindent=0pt]
    \item \editt{When the groundwater component of $\qsat$~\eqref{eq:qsat_dimensional} is much smaller than the overland component (\emph{e.g.} during intensive rainfalls),} the critical flow reached during extreme rainfalls can be approximated by
    \begin{equation}
        \label{eq:qsat}
        \qsat \approx r L_x \left(1 - \frac{K_s S_x L_z}{r_0 L_x}\right).
    \end{equation}
    
    Since the consecutive river flow rise is slow, the above estimate can be used as an approximation of the peak flow reached, assuming that the rainfall is long enough to reach the critical time, $\tsat$.
    
    \item Under this approximation, the critical point can be represented as a function of three parameters: rainfall intensity, catchment area (since the flow scales proportionally to both the hillslope width $L_x$ and catchment length $L_y$), and the $K_s S_x L_z/(r_0 L_x)$ factor, which is equal to the fraction of groundwater flow $K_s S_x L_z$ to the mean total flow $r_0 L_x$. This last parameter can be related to what is often referred to as the Base Flow Index, BFI~\citep[Sec. 3.1.2]{gustard1992low}.
    
    \item \edit{There are some similarities between this expression and other models used in hydrology. Equation~\eqref{eq:qsat} is a special case of the so-called rational method, which assumes that river flow is proportional to area and precipitation rate (see \citealt[chap. 1.5]{bedient2008hydrology}). The proportionality constant (runoff coefficient) here is identified as $1-\text{BFI}$.}
    
    \edit{The Base Flow Index appears in many statistical methods used in flood estimation, which, unlike our physically-based approach, are based on applying statistical methods such as linear regression to the available catchment data. A notable example is the Flood Estimation Handbook (FEH) flood estimation method by \cite{kjeldsen2008improving}. It assumes that the median of the annual maximum flow (QMED) scales as $\text{QMED}\propto 0.0460^{\text{BFIHOST}^2}$, where BFIHOST is a soil-based base flow index (BFI) estimator. Note that, similarly to \eqref{eq:qsat}, the predicted flow decreases with the base flow index, but in a nonlinear fashion. Interestingly, other catchment descriptors used in the FEH method include the catchment's area and precipitation, which, like base flow index, are also related to the maximum annual flow through nonlinear functions, selected to fit the available data.}
\end{enumerate}


\editt{Even though our 1D benchmark model is based on a series of simplifying assumptions (\emph{e.g.} a thin porous layer and pre-existing seepage zone), which are often not satisfied in real world catchments, its predictions seem to be reasonable in comparison with data based catchment models.} The connection between our simple scaling laws, shown above and derived analytically from a physical model, and statistical models such as the aforementioned FEH method, which are formulated in a completely different fashion, is intriguing. These results will be presented in a forthcoming work by the present authors\edit{, and can also be found in \cite{morawiecki2023phd}.}

\section{Conclusions}

\label{sec:conclusions}

\noindent The primary aim of our work has been to develop and analyse a rigorous benchmark scenario for coupled surface/subsurface flows in a typical catchment. We have achieved this goal by firstly characterising the typical parameter scales according to the available data on UK catchments (Part 1), formulating and computing the 3D model and its reduction (Part 2), and finally applying methods in asymptotic analysis to a reduced model valid for catchments dominated by overland dynamics (Part 3). 

In this last work, our analysis yields valuable scaling laws for the peak flows (see \cref{sec:hydrograph_features}), which precisely quantify the separation of time-scales observed in the hydrographs following an intense period of rain (see a distinct early- and late-time behaviour in \cref{fig:hydrograph_approximations}). In particular, we find that the early-time behaviour is governed by rainfall accumulation over a pre-existing seepage zone, followed by a slower flow rise in late-time. This latter stage is limited by the speed with which the rising groundwater increases the size of the seepage zone.

All approximations are in good agreement with hydrographs produced by the more complete 1D and 2D models, and allow accurate prediction of river peak flows over a wide range of catchment parameters (as long as the underlying assumptions are satisfied). \editt{However, different regimes not caputred by our model could be studied, including for example behaviour of catchments with no initial seepage ($\rho_0<1$), and late-time catchment behaviour in case of long continous rainfalls ($t=\Oh(1)$).}

\section{Discussion}
\label{sec:discussion}

\noindent Our investigations in these three parts have been limited to fairly elementary scenarios and geometries. However, our final results involving the derivation of analytical/asymptotic scaling laws with a clear underlying structure may serve as a valuable benchmark for other hillslope or catchment models. Currently, we observed that the benchmarking of coupled surface-subsurface catchment models has been limited to quantitative numerical comparisons, either with real-world observations or with the numerical output of other schemes (\emph{e.g.}~\citealt{maxwell2014surface}). While such studies allow practitioners to evaluate the given model's performance in specific conditions, they do not necessarily allow one to draw general conclusions about each model's limit of applicability. \edit{As a result, we have no guarantee that a given model will still perform well if applied in situations not captured in the training or validation data as demonstrated in many studies (\emph{e.g.} \citealt{klemevs1986operational} and \citealt{beven2019make}).}

\subsection{Applications of the benchmark to model intercomparisons}

\noindent Deriving asymptotic estimates for the hydrograph, as done in this work, opens another possibility: \textit{models can be compared at a more fundamental level}. For instance, we can check if the flows produced by various models scale with the different catchment properties in the expected fashion (and under the conditions we have specified). \edit{Detecting situations or limits where the predictions diverge can allow us to better understand the limitations of different models and shed light on how these models can be extended beyond their current limit of applicability. This idea is explored in our two forthcoming works \citep{paper_qmed, paper_RR}.}

As a particular example, statistical models used for flood estimation \edit{(such as the Flood Estimation Handbook (FEH) method, briefly discussed at the end of \cref{sec:hydrograph_features})} require the selection of empirical catchment descriptors for use in regression formulae to predict flood response \citep{kjeldsen2008improving}. \edit{In a forthcoming paper \citep{paper_qmed}, we show that the expression \eqref{eq:qsat} can be used to derive a simple expression for predicting peak monthly and annual river flows. This prediction turns out to be highly accurate when applied to real-world data. We then use this result to discuss the limitations of the existing statistical model from the Flood Estimation Handbook.}

Similarly, we can compare the analytical solutions developed within the physical benchmark model with the flow hydrographs generated via conceptual rainfall-runoff models (see models overview by \citealt{peel2020historical}). Although these models use continuous-time rainfall data to generate hydrographs, they are typically not based on the same fluid dynamical models of surface and subsurface flows. Consequently, they can be characterised by different scaling laws than the ones found in this paper. \edit{In our forthcoming work \citep{paper_RR}, we demonstrate this discrepancy by studying two models. The first is a Probability-Distributed Model (PDM) used by \textit{e.g.} the Environment Agency National Flood Forecasting System \citep{moore2007pdm}. The second is the Grid-to-Grid model by \cite{bell2007development}, used by the UK Centre for Ecology \& Hydrology to provide real-time flow predictions in the UK. In both cases, our simple benchmark scenario allows the identification of key differences between physical and aforementioned conceptual models.} 



\subsection{Extensions and generalisations of the benchmark scenario}

\noindent Another important line of inquiry is the generalisation of the analysis we have presented to situations that are more representative. Such extensions can involve the analysis of non-uniform rainfall, varying initial conditions to study the response to extended periods of drought\edit{, or catchments response to sudden drawdown or outlet water level. These asymptotic regimes have already been studied using the Boussinesq approximation (see \emph{e.g.} \citealt{mizumura2002drought} and \citealt{parlange2001sudden}). However, our coupled surface-subsurface approach could allow us to better understand the potential role of the seepage dynamics.} Then, analytical approximations of the drying process could be used to assess the assumptions of the conceptual rainfall-runoff models, especially since dry periods are sometimes used for a preliminary parameter calibration~\citep{lamb1999calibration}. Finally, we highlight the importance of multi-porosity regions in hydrological modelling; these lead to effects such as a preferential flow \citep{beven2013macropores}. We provide further details on potential extensions of this study in \citet[chap. 9]{morawiecki2023phd}.

\edit{Lastly, it would be interesting to obtain experimental validation of the studied regimes. There have been quite a few lab- and field-scale experimental studies, in which a constant rainfall was artificially generated. However, many of them (\emph{e.g.} \citealt{pauwels2018confirmation}) focus on systems limited to the groundwater flow only. There are some experimental studies in which rainfall we observed to yield seepage growth, \emph{e.g.} \cite{abdul1989field}, \cite{kollet2017integrated}, and \cite{scudeler2017examination}. However, in these experiments, the soil was initially dry, \emph{i.e.} there was no initial seepage, so they correspond only to the $\rho_0=0$ case. It would be interesting to conduct controlled experiments with more realistic settings, \emph{i.e.} with an already developed groundwater table, and compare the resultant hydrograph with our model predictions.}

\mbox{}\par
\noindent \textbf{Acknowledgements.} We thank Sean Longfield (Environmental Agency) for many useful interactions and for motivating this work via the 7th Integrative Think Tank hosted by the Statistical and Applied Mathematics CDT at Bath (SAMBa). We also thank Thomas Kjeldsen (Bath), Tristan Pryer (Bath), and Rob Lamb (Lancaster/JBA Trust) for insightful discussions. We are indebted to the reviewers and the JFM editorial team---their comments and suggestions were instrumental in the final development of this paper. Piotr Morawiecki is supported by a scholarship from the EPSRC Centre for Doctoral Training in Statistical Applied Mathematics at Bath (SAMBa), under the project EP/S022945/1.


\mbox{}\par
\noindent \textbf{Declaration of Interests.} The authors report no conflict of interest.

\bibliographystyle{plainnat}
\bibliography{bibliography}

\newpage
\appendix

\section{List of symbols}

\noindent For ease of reference, we provide a list of symbols in \cref{tab:list}.

{
\begin{tabular}{p{0.15\textwidth}p{0.11\textwidth}p{0.58\textwidth}}
    \footnotesize \\
    \toprule
    \textsc{group} & \textsc{symbol} & \textsc{description} \\
    \midrule
    Coordinates & $x$ & spatial coordinate along the hillslope \\
    & $t$ & time \\
    \midrule
    Variables & $H_g$ & groundwater depth \\
    & $h_s$ & surface water depth \\
    & $H$ & total depth ($H_g+h_s$) \\
    & $H_0$ & initial value of total depth \\
    & $Q_g$ & groundwater flux \\
    & $Q_s$ & overland flux \\
    \midrule
    Catchment & $L_z$ & thickness of the porous layer \\
    properties & $S_x$ & slope along the hillslope \\
    & $r$ & rainfall intensity \\
    & $K_s$ & hydraulic conductivity \\
    & $f$ & mean drainable porosity (typically a function of $x$) \\
    & $\theta_s, \theta_r$ & saturated and residual water content \\
    & $\alpha, n, m$ & Mualem-Van Genuchten model parameters \\
    & $T_0$ & characteristic timescale of groundwater flow \\
    & $n_s$ & Manning roughness coefficient \\
    & $k$ & exponent form the Manning's law (typically $k=5/3$) \\
    \midrule
    Dimensionless & $\sigma$ & porous layer thickness to elevation drop ratio \\
    constants & $\mu$ & overland to groundwater flux ratio \\
    & $\rho$ & precipitation rate to groundwater flux ratio \\
    & $\rho_0$ & mean precipitation rate to groundwater flux ratio \\
    & $\Delta\rho$ & difference between $\rho$ and $\rho_0$ \\
    & $\Pe$ & P\'eclet number for the overland flow \\
    & $a$ & saturated fraction of the hillslope \\
    & $a_0$ & leading-order approximation of $a$ for $\Pe\to\infty$ \\
    \midrule
    Characteristic &
    $T$ & rescaled time for the overland flow model \\
    curves & $\eta$ & rescaled surface water height for the overland flow model \\
    & $\tau$ & characteristic curve parameter \\
    & $s$ & parameter describing initial data \\
    & $T^*, \eta^*, Q_s^*$ & value of time, surface water height, and surface flow, when the characteristic curve reaches $x=0$ \\ 
    & $\tsat$ & critical time \\
    & $\qsat$ & critical flow \\
    \midrule
    Simulation & $N_x$ & spatial mesh resolution \\
    parameters & $N_t$ & number of time steps \\
    \bottomrule
\end{tabular}
\captionsetup{type=table}
\captionof{table}{List of symbols \label{tab:list}}
}

\newpage
\section{Relation between the 1D Boussinesq equation and the 2D Richards equation}

\label{app:Boussinesq_derivation}

\noindent In this appendix, we provide additional details on the derivation of the 1D model presented in this paper, in connection with the physical models presented in Part 2. We shall explain how the governing equations~\eqref{eq:groundwater_dimensional} and~\eqref{eq:overland_dimensional} relate to the two-dimensional Richards equation given in eqn (5.9a) from Part 2~\citep{paper2}. Some parts of the following are classical, and relate to the derivation of the Boussinesq equation [cf.~\cite{bear1987modeling} for details]; our new contribution is to consider the influence of the overland flow in the seepage zone on the Boussinesq equation and to couple this latter equation with its standard formulation for the remaining part of the domain.

In this study, we will assume that hillslope flow is predominantly two-dimensional in the $xz$ cross-section; this reduction from three dimensions to two dimensions is discussed in detail in Part 2. In addition, we consider the small aspect ratio limit, $L_z\ll L_x$, and hence $\beta_{zx} =L_z/L_x \to 0$. In the groundwater region, following (5.9a) from Part 2, the leading-order flow then satisfies 
\begin{equation}
    \label{eq:Richards_leading_order}
    \frac{\d\theta}{\d h}\bigg\rvert_{h=h_g'}\dt{h_g'} =
    \dzhat{} \left[K_r(h_g') \left(\dzhat{h_g'}+1\right)\right], \qquad 0 < \hat{z} < 1,
\end{equation}
with $\hat{z} = 0$ corresponding to the bottom of the aquifer and $\hat{z} = 1$ corresponding to the top surface. Note that for $\hat{z} < H$, the Richards equation becomes much simpler, since for saturated soil we have $\frac{\d\theta}{\d h}=0$ and $K_r(h)=1$. Solving \eqref{eq:Richards_leading_order} and imposing the no-flow boundary condition at the bottom of the aquifer $\hat{z}=0$, we obtain:
\begin{equation}
    \label{eq:hg_hydrostatic_profile}
    h_g'(\hat x,\hat z,t)=H'(\hat x,t)-\hat z, \qquad 0 < \hat{z} < H'(\hat{x}, t),
\end{equation}
which corresponds to a hydrostatic vertical profile of pressure. Note that based on the above solution, for regions of completely saturated soil, the two-dimensional function $h_g'(\hat x,\hat z,t)$ can be replaced by the one-dimensional indicator, $H'(\hat x,t)$. The curve $\hat{z} = H'(\hat{x}, t) \in (0, 1]$ corresponds to the groundwater table, which separates between saturated (where $h_g' > 0$) and unsaturated (where $h_g' < 0$) regions, if both coexist.

\begin{figure}
    \centering
    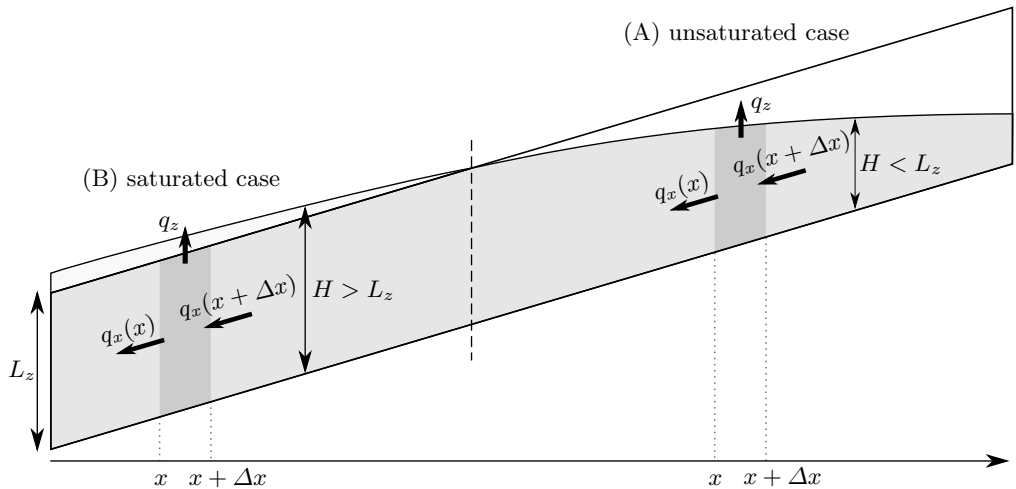
    \caption{Control volume (A) outside the seepage zone and (B) inside the seepage zone. One-way arrows represent flow in and out of the control volumes.}
    \label{fig:control_volume}
\end{figure}

We assume that the system is configured as shown in \cref{fig:control_volume}. Thus, it is divided into a Region B (fully saturated case), where $0 \leq \hat{x} \leq a(t)$ and the ground is entirely saturated, with $H = 1$. Similarly, we have Region A (unsaturated case) for $\hat{x} > a(t)$, where $H < 1$ and there is an unsaturated column where $H < \hat{z} < 1$. 

For ease of interpretation, we shall return, for the next few subsections, to dimensional quantities [related to the governing equation~\eqref{eq:dHdt_dimensional}]. Additionally, it is more convenient to use Richards equation expressed in the Cartesian coordinates $(x, z)$. In this coordinate system,~\eqref{eq:hg_hydrostatic_profile} becomes
\begin{equation}
    \label{eq:hg_hydrostatic_profile_xz}
    h_g(x,z,t) = S_x x + H(x,t) - z, \qquad \text{for $x\in [0,L_x]$, $z\in [0,L_z]$},
\end{equation}
where $L_z\hat{z}=z-S_x x$, $H=L_zH'$ and $h_g=L_zh_g'$. Following Darcy's law, we have $\mathbf{q}=K_s \nabla \left(h_g+z\right)$. After substituting~\eqref{eq:hg_hydrostatic_profile_xz}, we obtain the leading term of horizontal flow:
\begin{equation}
    \label{eq:app1_qx}
    q_x=-K_s\dx{} \left(h_g+z\right)=-K_s\left(S_x+\dx{H}\right).
\end{equation}
Note that the above point flux is independent of $z$. Multiplying by $H$, we thus see that the total (unsigned) flow along the hillslope is then given by
\begin{equation}
    \label{eq:dimensional_groundwater_flow_app}
    Q_g = H |q_x| = K_s H\dx{H} + K_s H S_x.
\end{equation}
Next, we shall consider the unsaturated part of the hillslope (Region A) and the seepage zone (Region B) separately since the latter requires adding an overland flow component.

\subsection{(Region A) Governing equation outside the seepage zone, \texorpdfstring{$H<1$}{}}
\label{app:derivation_unsaturated_equation}

\noindent Let us consider Region A, where there is a layer of unsaturated soil. Recall that the pressure $h_g$ is zero at the free surface of the groundwater, where $z=S_xx+H_g(x,t)$. This allows us to set the integration constant $H$ appearing in \eqref{eq:hg_hydrostatic_profile_xz} to $H(x,t)=H_g(x,t)$.

We consider the control volume, $V$, within the column horizontally bounded by $[x, x + \Delta x]$, as shown in \cref{fig:control_volume}. By conservation and the divergence theorem, it is argued that the total flux integral around the boundary, $\partial V$, is zero, and hence:
\begin{multline}
    0 = \oint_{\partial V} \nabla (h+z) \cdot \mathbf{n}\;\mathrm{d}l \\
    = - H(x)q_x(x) + H(x+\Delta x)q_x(x+\Delta x) +
        \int_x^{x+\Delta x} q_z\Bigr|_{z=S_x x + H(x)} \mathrm{d}x.
\end{multline}
Richards equation in the fully saturated groundwater region reduces to $\nabla\cdot\mathbf{q}=K_s\nabla^2\left(h+z\right)=0$ [as discussed following \eqref{eq:Richards_leading_order}]. Thus, in the $\Delta x\rightarrow 0$ limit, we can approximate the vertical flux at the top of the saturated zone as:
\begin{equation}
    \label{eq:groundwater_top_bc_flux}
    q_z^\mathrm{top}=q_z\Bigr|_{z=S_x x + H(x)}=-\dx{}\left(H q_x\right)=K_s\dx{}\left(HS_x+H\dx{H}\right),
\end{equation}
where we have used the expression for $q_x$ derived in~\eqref{eq:app1_qx}.

\edit{Now, let us consider the unsaturated zone above the groundwater table (dashed region in \cref{fig:control_volume}). The inflow from the top boundary is $r(t)$, and the outflow to groundwater is $q_z^\mathrm{top}$. If during a rainfall the inflow is greater than the outflow, the total volume of water in the soil column per surface area,
\begin{equation}
    \label{eq:soil_water}
    \mathcal{V}=\int_0^{L_z}\theta(h(z,t))\d z,
\end{equation}
increases, eventually leading to a rise of the groundwater table $H$ over time.}

\edit{In order to find the exact rate of change of $H(x,t)$, one should solve \eqref{eq:Richards_leading_order} and find $H(x,t)$ for which $h_g(x,z=H(x,t),t)=0$. Using the chain rule, we have
\begin{equation}
    \dt{}h_g(z=H(x,t),t)=\dz{h_g(x,z,t)}\Bigg|_{z=H(x,t)}\dt{H(x,t)}+\dt{h_g}\Bigg|_{z=H(x,t)}=0,
\end{equation}
from which the groundwater growth rate is
\begin{equation}
    \dt{H(x,t)}=-\left(\dz{h_g(x,z,t)}\right)^{-1}\dt{h_g}\Bigg|_{z=H(x,t)}.
\end{equation}
}

\edit{
However, rather than solving Richards equation in the unsaturated zone, a standard time-dependent Boussinesq-based approach to groundwater modelling is based on introducing a drainable porosity defined as
\begin{equation}
    f \equiv \dd{\mathcal{V}}{H}.
\end{equation}
In essence, the drainable porosity describes, for a given change in the height $H$, the corresponding change in the subsurface water volume $\mathcal{V}$ \eqref{eq:soil_water}. It is often assumed that the drainable porosity $f$ is a constant parameter characterising a given soil. However, as we argue in \cref{app:drainable_porosity}, this assumption is not consistent with the Richards-based approach, in which the growth of the volume $\mathcal{V}$ does not have to immediately lead to a rise in the groundwater volume. Therefore, in the same appendix, we introduce a mean drainable porosity $f(x)$, which does not fully represent the dynamics governed by the Richards equation but still allows us to exactly reproduce the time when the groundwater reaches the surface in the Richards-based approach.}

\edit{Since the volume of subsurface water changes as a result of precipitation $r(x, t)$ and inflow/outflow $q_z^\mathrm{top}$, we have 
\begin{equation}
    f(x)\dt{H}=\dd{\mathcal{V}}{t}=q_z^\mathrm{top}+r(x,t).
\end{equation}}
Substituting~\eqref{eq:groundwater_top_bc_flux} into the above expression gives us the governing equation for $H$:
\begin{equation}
    f(x)\dt{H}=K_s\dx{}\left(HS_x+H\dx{H}\right)+r(x,t).
\end{equation}
This equation is equivalent to~\eqref{eq:groundwater_dimensional} under the additional assumptions that infiltration is equal to precipitation --- as in Part 2 we ignore the effects of evapotranspiration, and additionally, we assume that no overland flow is generated unless the soil becomes fully saturated (\emph{i.e.} rainfall never exceeds soil infiltration capacity, $r(x,t)\leq K_s$).

\subsection{(Region B) Governing equation for seepage zone \texorpdfstring{$H\geq 1$}{}}
\label{app:derivation_saturated_equation}

\noindent In the case of the seepage zone, the groundwater height is fixed at $H_g=L_z$. However, we need to determine the additional surface water height, $h_s(x,t)$, which is given by solving the Saint Venant equation [eqn (3.3) from Part 2]:
\begin{equation}
    \label{eq:St_Venant_overland}
    \dt{h_s}=\dx{}\Bigl[Q_s(h_s)\Bigr] + r(x, t) - I(x, t),
\end{equation}
where the surface flow, $Q_s(h_s)$, is given by the Manning's equation~\eqref{eq:dimensional_Mannings}, and $I$ is the infiltration rate, equal to the negated vertical flow, $-q_z$, at ground level.

To determine the vertical flow, $q_z$, we perform a similar conservation argument as in \cref{app:derivation_unsaturated_equation}. Firstly, at the interface between the subsurface and surface flows, we set continuity of pressure and flow. The first condition allows us to specify the pressure head, $h_g$, at the surface:
\begin{equation}
h_g(x, z, t) = h_s(x, t) \quad \text{at $z = S_x x + L_z$}.
\end{equation}
Consequently, we set the integration constant $H$, appearing in \eqref{eq:hg_hydrostatic_profile_xz}, to $H(x,t)=L_z+h_s(x,t)$. Substituting it into~\eqref{eq:app1_qx}, we obtain:
\begin{equation}
    \label{eq:app1_qx_saturated_case}
    q_x=-K_s\left(S_x+\dx{h_s}\right) \qquad \text{for }0 < z < L_z.
\end{equation}

Now, let us consider the control volume of groundwater contained in the vertical column $[x, x + \Delta x]$, as illustrated in \cref{fig:control_volume} in Region B. In this case, we have $H_g=L_z$, and the mass balance equation~\eqref{eq:groundwater_top_bc_flux} yields
\begin{equation}
    \label{eq:qz}
    q_z\Bigr|_{z=S_x x + L_z}=-L_z\dx{q_x} = \dx{}\left(K_sL_zS_x+K_sL_z\dx{h_s}\right) = \dx{}\left(K_sL_z\dx{h_s}\right),
\end{equation}
after substituting the expression \eqref{eq:app1_qx_saturated_case} for the groundwater flow, $q_x$. Thus, we have the required expression for the infiltration rate:
\begin{equation}
    \label{eq:infiltration}
    I(x, t) =-q_z\Bigr|_{z=S_x x + L_z} = -\dx{}\left(K_sL_z\dx{h_s}\right).
\end{equation}
Note that the infiltration is positive if surface water infiltrates into the soil or negative if groundwater emerges to the surface.

By substituting $Q_s$ from~\eqref{eq:dimensional_Mannings} and infiltration $I$ from~\eqref{eq:infiltration} into \eqref{eq:St_Venant_overland}, we obtain:
\begin{equation}
    \dt{h_s}=\dx{}\left(K_sL_z\dx{h_s}+\frac{\sqrt{S_x}}{n_s}  h_s^k\right)+r(x,t).
\end{equation}
Note that in the seepage zone, $H=H_g+h_s$, where $H_g=L_z$ is constant. Hence, this equation is equivalent to the governing equation~\eqref{eq:overland_dimensional}. Thus, we have derived both cases of the governing equation~\eqref{eq:dHdt_dimensional}, forming the basis of this work.

\section{Mean drainable porosity function, \texorpdfstring{$f(x)$}{}}
\label{app:drainable_porosity}

\subsection{On the mean drainable porosity}
\label{app:vertical_flow}

\edit{In \cref{app:Boussinesq_derivation}, we defined the drainable porosity $f$ as the volumetric change in the subsurface volume for a given change in the groundwater height, $\de{\mathcal{V}}/\de{H}$. However, this approach is not fully consistent with the solution for the Richards equation for subsurface flow.}

Typically, in scenarios where the precipitation increases to a constant value $r>1$, a characteristic wetting front is observed. This front can be seen in fig.~8d in Part 2, and its propagation is discussed in detail by~\citealt{caputo2008front}. The front moves downward towards the groundwater table, changing soil saturation and eventually leading to the rise of the groundwater table. This behaviour is not captured by a standard time-dependent Boussinesq equation or our one-dimensional model.

However, our analysis in sections~\ref{sec:groundwater_rising} and~\ref{sec:method_of_characteristics} shows that the most important mechanism by which groundwater contributes to the peak flow generation is by extending the seepage zone. Moreover, in~\cref{app:groundwater_outer_derivation}, we show that horizontal groundwater flow is negligibly slow and does not impact the speed at which the groundwater is rising over a short timescale characterising a typical rainfall. This means that the groundwater becomes saturated when rainwater fills the available drainable volume $v_h$, \emph{i.e.} after time $t=v_h/r$. The same time is predicted by the 1D model by setting an $x$-dependent mean drainable porosity $f(x)$, defined as
\begin{equation} \label{eq:mean_porosity_app}
    f(x)=\frac{v_H(x)}{D(x)},
\end{equation}
where $D(x)$ is the thickness of the unsaturated soil layer. The above choice of mean drainable porosity is the one that we use throughout this work.

To summarise, even though setting a time-independent porosity and constant groundwater recharge does not capture the delay required for the wetting front to reach the groundwater, it allows for the correct prediction of the soil critical time and the resultant peak flows observed in this work. In this model, note that $H(x,t)$ does not represent the exact thickness of the saturated zone that forms groundwater but corresponds to the amount of water absorbed by the soil, even if it has not yet reached the saturated zone. However, this difference in interpretation does not seem to affect the main results obtained in this paper. Formulating this complex system in terms of a 1D partial differential equation facilitates the analysis and computation.

\subsection{Computation of mean drainable porosity}
\label{sec:mean_drainable_porosity}

\noindent Here, we formally derive an expression for the mean porosity given the 1D model parameters for the considered benchmark scenario. Firstly, following the Richards equation, we find the pressure head $h_g$ profile above the groundwater table and use it to evaluate the drainable volume for a column of soil of height $D(x)$ above the groundwater table.

We note that in the considered case of $\beta_{zx} = L_z/L_x \ll 1$, the leading solution of the two-dimensional Richards equation involves only vertical flow along the $\hat{z}$ axis, as given by~\eqref{eq:Richards_leading_order}. In the scenario considered in this paper, we assume that the system is initially in a steady state for a constant rainfall $r_0$. Under these assumptions, we integrate the time-independent version of~\eqref{eq:Richards_leading_order} to form a first-order nonlinear ODE for the pressure head $h_g(\hat z)$:
\begin{equation}
    \label{eq:Richards_steady_state}
    K_r(h_g) \left(\dd{h_g}{\hat{z}}+1\right) = \frac{r_0}{K_s},
\end{equation}
where the constant of integration on the right-hand side has been chosen to match the dimensionless infiltration $r_0/K_s$, passing through the surface. Let us consider a column of the soil above a groundwater table, and we take $h_g\Big|_{\hat{z}=0}=0$. Given the pressure head, we can find the saturation $\theta(h_g(\hat z))$ following the Mualem-Van Genuchten model:
\begin{equation}
    \theta(h_g) =
    \begin{cases} 
        \theta_r+\frac{\theta_s-\theta_r}{\left(1+\left(\alpha h_g\right)^n\right)^m} & h_g<0 \\
        \theta_s & h_g\ge 0
    \end{cases},
    \label{eq:GMtheta}
\end{equation}
\edit{where $\theta_s$ and $\theta_r$ are the saturated and residual water content, while $\alpha$, $n$, and $m=1-\frac{1}{n}$ are other Mualem-Van Genuchten parameters characterising a given soil.} We can use \eqref{eq:GMtheta} to compute the drainable volume and resulting mean drainable porosity~\eqref{eq:mean_porosity_app}:
\begin{equation}
    \label{eq:f_numerical}
    f(x) = \frac{v_H(x)}{D(x)} = \frac{1}{D(x)} \int_{0}^{D(x)} \Big[\theta_s-\theta(h_g(\hat z))\Big] d\hat z.
\end{equation}
Solving~\eqref{eq:Richards_steady_state} and then numerically integrating~\eqref{eq:f_numerical} allows us to calculate $f$.

\subsection{Analytical approximations of mean drainable porosity}

\begin{figure}
    \centering
    \begin{preview}
    \begin{tikzpicture}[font=\small]

\tikzset{legendstyle/.style={anchor=west,inner sep=0pt, outer sep=0pt,text width=180, scale=0.85}}
\definecolor{color1}{RGB}{168,221,181}
\definecolor{color2}{RGB}{123,204,196}
\definecolor{color3}{RGB}{67,162,202}

\begin{axis}[
    width=0.45\linewidth,
    height=0.4\linewidth,
    xmin={0},
    xmax={1},
    ymin=-0.9,
    ymax=0,
    xlabel={},
    ylabel={hydraulic head, $h_g\left(\hat z\right)$},
    xticklabels={,,},
    legend pos=south east,
    at={(0,0)},
    ]

\addplot [color3, thick] table [mark=none, x=z, y=h, col sep=comma] {DATA/vertical_profile.dat};
\addlegendentry[anchor=west,scale=0.85]{full solution}

\addplot [color2, dashed, thick] table [mark=none, x=z, y=h_approx, col sep=comma] {DATA/vertical_profile.dat};
\addlegendentry[anchor=west,scale=0.85]{approximation~\eqref{eq:hg_linear_approximation}}

\end{axis}

\begin{axis}[
    width=0.45\linewidth,
    height=0.4\linewidth,
    xmin={0},
    xmax={1},
    ymin={0.35},
    ymax={0.495},
    xlabel={height above the groundwater table, $\hat z$},
    ylabel={soil saturation, $\theta\left(\hat z\right)$},
    at={(0,-0.29\linewidth)},
    ]

\addplot [color3, thick] table [mark=none, x=z, y=theta, col sep=comma] {DATA/vertical_profile.dat};

\addplot [color2, dashed, thick] table [mark=none, x=z, y=theta_approx, col sep=comma] {DATA/vertical_profile.dat};

\addplot[mark=none, gray, dotted, thick] coordinates {(0, 0.488) (1, 0.488)}; 
\node[gray, right] at (axis cs:0.6,0.48){$\theta_s=0.488$};

\end{axis}

\begin{axis}[
    width=0.5\linewidth,
    height=0.683\linewidth,
    xmin={1e-3},
    xmax={1},
    ymin={1e-5},
    ymax={1},
    xmode=log,
    ymode=log,
    xlabel={groundwater depth below the surface, $D$},
    ylabel={mean porosity, $f(D)$},
    legend pos=north west,
    at={(0.47\linewidth,-0.283\linewidth)},
    ],

\addplot [color3, thick] table [mark=none, x=D, y=f, col sep=comma] {DATA/porosity_approximations.dat};
\addlegendentry[anchor=west,scale=0.85]{full solution}
\addplot [color2, dashed, thick] table [mark=none, x=D, y=f1, col sep=comma] {DATA/porosity_approximations.dat};
\addlegendentry[anchor=west,scale=0.85]{approximation~\eqref{eq:f_approx_1}}
\addplot [color1, dashdotted, thick] table [mark=none, x=D, y=f2, col sep=comma] {DATA/porosity_approximations.dat};
\addlegendentry[anchor=west,scale=0.85]{approximation~\eqref{eq:f_approx_2}}

\addplot[mark=none, gray, dotted, thick] coordinates {(0.002, 1.5352e-4) (0.01, 1.0422e-3)}; 
\node[gray, right] at (axis cs:2e-3,6e-4){$D^n$};
\end{axis}

\end{tikzpicture}
    \end{preview}
    
    \caption{On the left: Initial vertical profile of the pressure head $h_g$ and corresponding saturation $\theta$ in a column of soil. On the right: Dependence of mean porosity $f=\frac{v_h}{D}$ on the depth of the groundwater below the surface. As shown, approximations~\eqref{eq:hg_linear_approximation}-\eqref{eq:f_approx_2} accurately describe soil properties close to the groundwater table. We used MvG parameter values from the previous part of our work, \emph{i.e.} $\alpha=3.367\mathrm{m^{-1}}$, $\theta_s=0.388$, $\theta_R=0.115$, and $n=1.282$.}
    \label{fig:vertical_profile}
\end{figure}
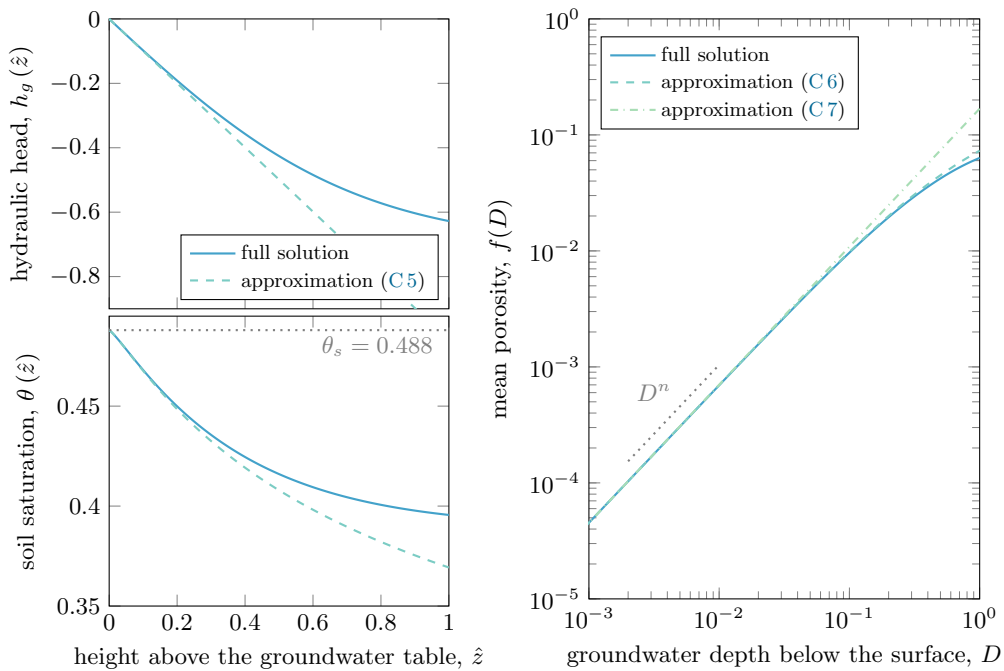

\noindent As an alternative to the integral expression above, let us develop an approximation for $v_H$, based on the assumption that the groundwater table is located near the land surface. Since for $h\rightarrow 0$, $K_r(h)\rightarrow 1$, the leading-order solution for~\eqref{eq:Richards_steady_state} around $\hat z=0$ satisfies
\begin{equation}
    \label{eq:hg_linear_approximation}
    h_g(\hat z) \sim \left(\frac{r_0}{K_s} - 1\right) \hat z.
\end{equation}
Then, integrating~\eqref{eq:f_numerical} with the MvG model~\eqref{eq:GMtheta} gives:
\begin{equation}
    \label{eq:f_approx_1}
    f_1(x) = \left(\theta_s-\theta_r\right)\left[{}_2F_1\left(m,\frac{1}{n};1+\frac{1}{n};\left(-a\left(\frac{r_0}{K_s} - 1\right)D(x)\right)^n\right)-1\right],
\end{equation}
where ${}_2F_1$ is a hypergeometric function. One can also find the leading-order approximation of~\eqref{eq:f_approx_1} for $D \ll 1$, which yields:
\begin{equation}
    \label{eq:f_approx_2}
    f_2(x) = \frac{m}{n+1}\left(\theta_s-\theta_r\right)\left[-\left(\frac{r_0}{K_s} - 1\right) \alpha D(x)\right]^n.
\end{equation}
However, the approximation~\eqref{eq:f_approx_1} is more accurate. Functions~\eqref{eq:f_approx_1} and~\eqref{eq:f_approx_2} are compared with the full numerical solution in \cref{fig:vertical_profile}.

\section{Derivation of steady-state overland flow for $\mu \to \infty$}
\label{app:steady_state}

\noindent The steady state for the seepage zone is given by equation~\eqref{eq:Hs_steady_state_integrated}:
\begin{equation}
    \label{eq:steady_state_deriv_0}
    1+\sigma\dx{h_0}+\mu h_0^k=\rho_0(1-x),
\end{equation}
where we used $h_0(x)=H(0)-1$ to represent the surface water height.

As we have noted previously, our regime of interest is where $\mu$ is large. It is convenient to re-scale via $x = aX$, where $a$ was defined as the size of the seepage zone, so that the domain is fixed in $X\in[0, 1]$. In the limit $\mu \to \infty$, we anticipate from dominant balance that the overland flow is small, $h_0 \to 0$, and hence we re-scale $h_0 = \mu^{-1/k}g(X)$ in order to balance the third term with the right-hand side of \eqref{eq:steady_state_deriv_0}. With the boundary conditions~(\ref{eq:bc_dimless}a,c), this rescaling gives
\begin{subequations}
    \begin{gather}
        \frac{\Pe ^{-1}}{a}\frac{\partial g}{\partial X}+g^k=\rho_0(1-aX) - 1, \label{eq:steady_state_deriv_1} \\
        g'(0) = 0 \quad \text{and} \quad g(1) = 0. \label{eq:steady_state_deriv_1_bcs}
    \end{gather}
\end{subequations}
\noindent In \eqref{eq:steady_state_deriv_1}, for convenience, we have defined the small parameter
\[
\Pe ^{-1} = \frac{\sigma}{\mu^{1/k}} \ll 1
\]
which, according to \cref{tab:typical_parameters}, is approximately $\Pe ^{-1} \approx 10^{-5}$. As we shall see, the asymptotic analysis as $\Pe ^{-1} \to 0$ involves the analysis of an outer solution, where $0 \leq X < 1$, and an inner solution with a boundary layer, where $X \to 1$. 

In the outer region, we expand the solution and contact line as
\begin{equation} \label{eq:gser_aser}
    g(X) = g_0(X) + \Pe ^{-1} g_1(X) + \ldots \qquad \text{and} \qquad
    a = a_0 + \Pe ^{-\gamma} a_1 + \ldots,
\end{equation}
where $\gamma > 0$ and is to be determined later. Then we develop the following leading-order approximation of the outer surface water height via \eqref{eq:steady_state_deriv_1}, 
\begin{equation}
    \label{eq:steady_state_deriv_2}
    g(X) \sim g_0(X) = \left[\rho_0(1-a_0 X)-1\right]^{1/k}.
\end{equation}
The above outer expansion exhibits an infinite gradient, possibly before the contact line is reached, at the point $X = \left(1 - \rho_0^{-1}\right)/a_0$. However, we show below that $a_0$ is such that this point corresponds to $X = 1$.

In order to develop the solution within the boundary layer, we consider re-scaling
\[
g(X) = \Pe ^{-\beta} \hat{G}(\hat{X}) \quad \text{and} \quad
X = 1 - \Pe ^{-\alpha} \hat{X}, 
\]
with $\alpha$, $\beta$, and $\gamma$ from \eqref{eq:gser_aser} to be determined. Applying these transformations to~\eqref{eq:steady_state_deriv_1} gives:
\begin{multline}
    \label{eq:steady_state_deriv_3}
    \Pe ^{-(1+\beta-\alpha)} \left[-\frac{1}{a_0}\frac{\partial \hat{G}}{\partial \hat{X}}\right] +\Pe ^{-k\beta} \biggl[ \hat{G}^k\biggr] = \\ \biggl[\rho_0(1-a_0) - 1\biggr] + \Pe ^{-\alpha} \biggl[\rho_0 a_0  \hat{X}\biggr] - \Pe ^{-\gamma}\biggl[\rho_0 a_1\biggr]  + o(\Pe ^{-\alpha})
\end{multline}
The balance of $\Pe $ terms is achieved for $1+\beta-\alpha=k\beta=\alpha=\gamma$, which corresponds to $\alpha = \gamma = k/(2k-1)$ and $\beta=1/(2k-1)$. The equation~\eqref{eq:steady_state_deriv_3} for the leading $\mathcal{O}(1)$ term is $\rho_0(1-a_0) - 1 = 0$, and hence this approximation gives us an estimate for the leading-order contact line position,
\begin{equation}
a\sim a_0=1-\frac{1}{\rho_0}.
\end{equation}
Indeed, this equation confirms that the leading-order outer solution, \eqref{eq:steady_state_deriv_2}, predicts an infinite gradient as $X \to 1$. In order to derive an approximation for the gradient of the surface water height near the contact line, we must proceed to the next order in the inner region.

The equation~\eqref{eq:steady_state_deriv_3} at $\mathcal{O}(\Pe^{-\frac{k}{2k-1}})$ yields
\begin{subequations}
\begin{equation}
    \label{eq:bvp_for_a1}
    -\frac{1}{a_0}\frac{\partial \hat{G}}{\partial \hat{X}}+\hat{G}^k= \rho_0 a_0 \hat{X} - \rho_0 a_1,
\end{equation}
along with the two boundary conditions:
\begin{equation}
    \hat{G}(0) = 0 \qquad \text{and} \quad
    \hat{G}(\hat{X}\rightarrow\infty) \sim (\rho_0-1)^{1/k} \hat{X}^{1/k}.
\end{equation}
\end{subequations}
The second boundary condition above corresponds to the inner limit of the outer solution~\eqref{eq:steady_state_deriv_2}. The above first-order ODE, together with the two boundary conditions, forms a boundary value problem with eigenvalue $a_1$. It cannot be solved analytically. In any case, if desired, the above problem can be computed numerically, and it would yield the correction to the contact line, $a_1 = a_1(\rho_0)$.

Notice finally that the gradient of $h_0$ at $x=a$ can now be estimated directly from equation~\eqref{eq:steady_state_deriv_0}. Since $h_0(a)=0$, we have, using the expansion \eqref{eq:gser_aser},
\begin{equation}
    \dx{h_0}\Big\rvert_{x=a} = \frac{\rho_0(1-a) - 1}{\sigma} \sim \frac{\rho_0(1-a_0) - 1 - \rho_0 a_1\Pe ^{-\gamma}}{\sigma} = - \frac{\rho_0 a_1}{\sigma}\Pe ^{-\gamma}.
\end{equation}

\begin{figure}
    \centering
    \includegraphics{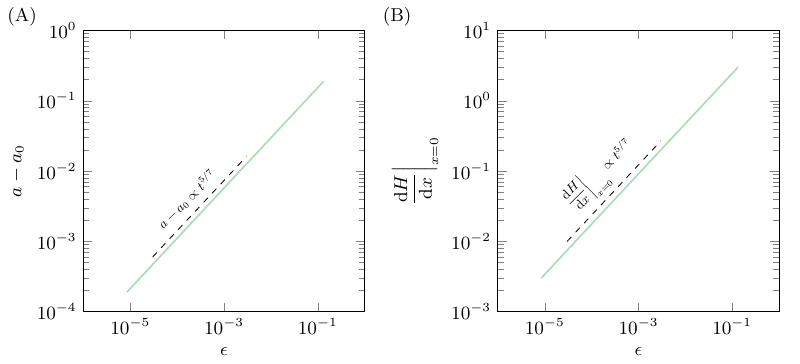}
    \caption{(A) The size of the seepage zone relative to its first-order approximation $a-a_0$. (B) The gradient of $H$ at this point. The results were obtained by solving eqn~\eqref{eq:steady_state_deriv_0}. The fitted power law is consistent with the theoretical exponent $\gamma=5/7$. }
    \label{fig:eps_dependence}
\end{figure}

For Manning's law, $k= 5/3$, so $\gamma = k/(2k-1) = 5/7$. Hence, we expect the gradient to be $\de{h_0}/\de{x} = \Oh(\Pe ^{-5/7})$ while the contact line position equally satisfies $a - a_0 = \Oh(\Pe ^{-5/7})$. These two scaling laws are confirmed by solving the boundary value problem given by equation~\eqref{eq:steady_state_deriv_1} numerically and transforming it back to the original variables (see \cref{fig:eps_dependence}).

\section{Derivation of the initial groundwater table for $\sigma\rightarrow 0$}
\label{app:H0_derivation}

\noindent Here, we find the leading-order solution for the ODE~\eqref{eq:H0_ode} under the limit $\epsilon=\sigma\rightarrow 0$:
\begin{equation}
    \label{eq:H0_ode_with_eps}
     \epsilon \dd{H_0}{x} = \frac{\rho_0(1-x)}{H_0(x)} - 1, \qquad \text{for } x\in[a_0,1],
\end{equation}
with a boundary condition $H_0(a_0)=1$, and where $a(t) \sim a_0$ is the leading-order contact line position as $\Pe \to \infty$. Expanding $H_0$ in powers of $\ep$, we obtain the approximation:
\begin{equation}
    \label{eq:outer_expansion_H0}
    H_0(x)= \rho_0(1-x) +\rho_0 \epsilon +\Oh(\epsilon^2).
\end{equation}
This solution does not satisfy the boundary condition at $x=a_0$. For $x\to a_0$, we develop an inner expansion by re-scaling $x=a_0+\epsilon X$ and $H_0(x) = g(X)$. The ODE now becomes:
\begin{equation}
    \dd{g}{X} = \frac{\rho_0(1-a_0-\epsilon X)}{g(X)} - 1, \qquad \text{for } X>0,
\end{equation}
with $g(0)=1$. Solving now to the first two orders, we have:
\begin{equation}
    \label{eq:inner_expansion_H0}
    g(X)=1+\epsilon \rho_0(1-e^{-X}-X) +\Oh(\epsilon^2),
\end{equation}
where we have used $a_0 = 1-1/\rho_0$ from \eqref{eq:a0}.

The composite solution is obtained by adding the outer and inner asymptotic expansions from  \eqref{eq:outer_expansion_H0} and \eqref{eq:inner_expansion_H0}, respectively, and subtracting their common part. The final result, to two orders of accuracy, is the following approximation:
\begin{equation}
    H_0(x)=\rho_0\left(1-x+\epsilon-\epsilon e^{-\frac{x-a_0}{\epsilon}}\right) + \Oh(\epsilon^2).
\end{equation}
After replacing $\epsilon$ with $\sigma$, we obtain the  solution~\eqref{eq:H0_matched_solution} stated in the main text.

\section{Derivation of the time-dependent groundwater solution for $t \to 0$ and $\rho \to \infty$}
\label{app:groundwater_outer_derivation}

\noindent We provide additional details for the early-time analysis of \cref{sec:groundwater_rising}, particularly in connection with the boundary-layer asymptotics. Consider equation~\eqref{eq:dHdt_unsaturated_zone} outside the seepage zone ($H<1$):
\begin{equation}
    \label{eq:dHdt_app}
    f(x) \dt{H} = \dx{}\left(\sigma H\dx{H} + H\right) + \rho.
\end{equation}
As noted in \S{\ref{sec:groundwater_rising}}, we consider the initial condition, $H(x, t = 0) = H_0(x; \, \rho_0)$, to be the steady-state response of the system to a precipitation rate, $\rho_0$. That is, the initial groundwater solution satisfies~\eqref{eq:H0_ode}:
\begin{equation}
    \left(\sigma H_0 H_0' + H_0\right)' + \rho_0 = 0, 
\end{equation}
subject to the boundary conditions (\ref{eq:bc_dimless}b,d) in \S{\ref{sec:steady_state}}, \emph{i.e.}:
\begin{equation}
H(x=a(t))=1 \quad\text{and}\quad \left(\sigma H\dx{H} + H\right) \Bigg\vert_{x=1} = 0.
\end{equation}

Let us consider the short-time behaviour of the time-dependent equation. Let $t=\epsilon t'$, $H(x,t)=H_0(x) + \epsilon H'(x,t)$, where $\epsilon\ll 1$. Then, \eqref{eq:dHdt_app} becomes:
\begin{equation}
    \label{eq:dHgdt_outer}
    f(x) \frac{\partial H'}{\partial t'} = \rho - \rho_0 + \epsilon \dx{}\left(\sigma H_0\dx{H'} + \sigma H'\dx{H_0} + H'\right) + \mathcal{O}(\epsilon^2).
\end{equation}
Therefore, the first two leading terms of the small-time groundwater solution are:
\begin{equation}
    \label{eq:Hg_outer}
    H(x,t) = H_0(x; \, \rho_0) + \left[\frac{\Delta \rho}{f(x)} \right]t + \mathcal{O}(t^2), \qquad \text{as $t \to 0$ with $x > a(t)$},
\end{equation}
where $\Delta \rho \equiv \rho-\rho_0$. Thus, we see that \eqref{eq:Hg_outer} predicts that the groundwater height increases linearly with time by an amount proportional to the sudden impulse of rain (or rather, its difference $\Delta \rho$). 

However, the above asymptotic solution assumes that $x - a(t) = \Oh(1)$, and indeed it fails to account for the fact that $H = 1$ when $x = a(t)$. We must consider it to be an \textit{outer} solution, valid away from $x = a(t)$. A comparison between the outer approximation \eqref{eq:Hg_outer} and the full solution is shown in \cref{fig:a_vs_t}(a). In the case of the fully coupled surface-subsurface flow, the solution of the PDE diverges from the outer solution~\eqref{eq:Hg_outer} within a small boundary layer around the seepage front.

\begin{figure}
    \centering
    \includegraphics{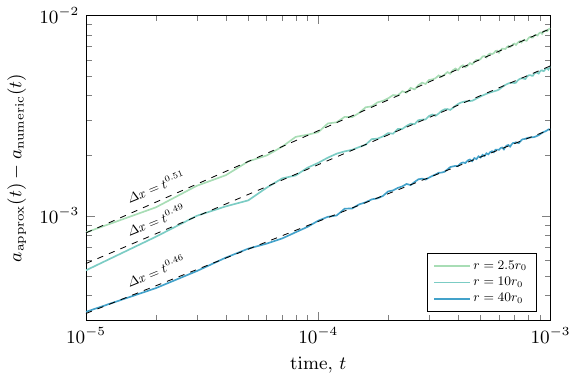}
    \caption{Difference between the location of the seepage front obtained by solving PDE~\eqref{eq:dHdt_dimless} and obtained using the leading-term outer solution~\eqref{eq:Hg_outer} for rainfalls with three different precipitation rates.}
    \label{fig:delta_x_vs_t}
\end{figure}

The size of the boundary layer tends to zero as $\rho \to \infty$. This is tested as follows. Firstly, we compute the full PDE model~\eqref{eq:dHdt_dimless} and find the place where $H(x_\mathrm{numeric})=1$. Then, we estimate the boundary layer thickness by finding the difference between $x_\mathrm{numeric}$ and $x_\mathrm{approx}=1-1/\rho$, which is the leading-order (outer) approximation of $a(t)$. As \cref{fig:delta_x_vs_t} demonstrates, it depends both on time $t$ and precipitation represented by $\rho$.

The above is confirmed via a dominant balance. Let $x=a+\delta x'$. Then equation~\eqref{eq:dHgdt_outer} becomes:
\begin{equation}
    \label{eq:dHgdt_inner}
    f(x) \frac{\partial H'}{\partial t'} = \rho - \rho_0 + \frac{\epsilon}{\delta^2} \dxp{}\left(\sigma H_0\dxp{H'} + \sigma H'\dxp{H_0} + \delta H'\right) + \Oh\left(\epsilon^2\right)
\end{equation}

The diffusion terms become significant and balance the precipitation $\rho$ when $\epsilon/\delta^2 \sim \Delta\rho$. Then the thickness of the boundary layer is on the order of $\delta=\epsilon^{1/2}\Delta\rho^{-1/2}$, \emph{i.e.} it increases proportionally to $t^{1/2}$ and $\Delta\rho^{-1/2}$. These trends are confirmed in \cref{fig:delta_x_vs_t}.

\edit{
\section{Derivation of the explicit solution for \texorpdfstring{$Q(t)$}{}}
\label{app:explicit_form}
In this appendix, we show that the implicit solution \eqref{eq:Q_implicit_1} for late-time (\emph{i.e.} for $Q_s^* > \rho a_0$),
\begin{equation}
    \label{eq:app_implicit}
    t^*(Q_s^*) = \frac{1}{\rho}\left(Q_s^*\right)^{1/k}+\mu^{1/k} \mathcal{T}\left(\frac{Q_s^*}{\rho}\right),
\end{equation}
can be written in an explicit form given by equations (\ref{eq:qs_explicit_group}a-c), as long as the increase of flow after the critical time is negligibly small compared to the to the critical flow value $\qsat=\rho a_0$. This assumption is motivated by the observation that the dynamics of the flow increase after reaching critical flow slows down significantly.
}

\edit{
Let us consider the flow $Q_s^*$ near its critical value, $Q_s^* = \rho a_0 + \epsilon Q'$, and consider the asymptotic limit $\epsilon\rightarrow 0$. Then \eqref{eq:app_implicit} becomes:
\begin{equation}
    t^* = \frac{1}{\rho}\left(\rho a_0 + \epsilon Q'\right)^{1/k}+\mu^{1/k} \mathcal{T}\left(Q_s^*\right).
\end{equation}
Therefore, we have:
\begin{equation}
    t^* = \underbrace{\frac{1}{\rho}\left(\rho a_0\right)^{1/k}}_{\tsat}+\mu^{1/k} \mathcal{T}\left(Q_s^*\right)+\Oh(\epsilon).
\end{equation}
By neglecting the $\Oh(\epsilon)$ term, we get:
\begin{equation}
    t^* - \tsat = \mu^{1/k} \mathcal{T}\left(Q_s^*\right),
\end{equation}
which allows us to obtain an explicit equation for $Q_s^*$:
\begin{equation}
    Q_s^* = \mathcal{T}^{-1}\left(\mu^{-1/k}\left(t^*-\tsat\right)\right)
\end{equation}
}

\edit{
Now we need to find an inverse function for $t=\mathcal{T}\Big(a(t)\Big)$, \emph{i.e.} a function that provides the location of the seepage front for a given time. Function $\mathcal{T}$ is defined by \eqref{eq:front_equation} as
\begin{equation}
    \label{eq:tau}
    \mathcal{T}(x=a(t)) \sim \frac{f(x=a(t))}{\rho - \rho_0} \Big(1 - H_0\big(x=a(t)\big)\Big).
\end{equation}
We use the approximation of $f(x)$ given by \eqref{eq:f_approx_2}, and the approximation of $H(x)$ given by \eqref{eq:H0_matched_solution},
\begin{subequations}
\begin{equation}
\label{eq:app_f}
    f(x)=CD(x)^n\text{,  where } C=\frac{m}{n+1}\left(\theta_s-\theta_r\right)\left[-\left(\frac{r_0}{K_s} - 1\right) \alpha\right]^n,
\end{equation}
\begin{equation}
    \label{eq:D}
    D(x)=1-H(x)=1-\rho_0\left(1-x+\sigma-\sigma e^{-\frac{x-a_0}{\sigma}}\right).
\end{equation}
\end{subequations}
By substituting \eqref{eq:app_f} to~\eqref{eq:tau} we get:
\begin{equation}
    \mathcal{T}(x=a(t)) = \frac{1}{\rho - \rho_0} C D(x=a(t))^{n+1}.
\end{equation}
Now, we solve this equation for $D(x=a(t))$:
\begin{equation}
    \left[C^{-1}\left(\rho - \rho_0\right)\mathcal{T}(x=a(t))\right]^\frac{1}{n+1} = D(x=a(t)).
\end{equation}
After substituting \eqref{eq:D} we get:
\begin{equation}
    \rho s(t) = A\rho t^\frac{1}{n+1} = 1-\rho_0\left(1-x+\sigma-\sigma e^{-\frac{x-a_0}{\sigma}}\right),
\end{equation}
where we introduced $A=\left[C^{-1}\left(\rho - \rho_0\right)\right]^\frac{1}{n+1}$ and $t=\mathcal{T}(x=a(t))$ to shorten the notation. The solution of this equation for $x$ is:
\begin{equation}
    x = a(t) = s(t)+\sigma+\underbrace{1-\frac{1}{\rho_0}}_{a_0}+\sigma W_0\left(-e^{-1-s(t)/\sigma}\right),
\end{equation}
where $W_0(\cdot)$ is the Lambert W function. Therefore, the solution can be written as:
\begin{equation}
    a(t)=\underbrace{a_0}_{\text{term 1}}+\underbrace{s(t)}_{\text{term 2}}+\underbrace{\sigma\left[1+W_0\left(-\mathrm{e}^{-1-s(t)/\sigma}\right)\right]}_{\text{term 3}}.
\end{equation}
This concludes the derivation of the explicit solution \eqref{eq:qs_explicit_group}. The accuracy of this approximation is demonstrated in \cref{fig:hydrograph_approximations}.
}

\end{document}